\documentclass[12pt]{iopart}

\usepackage[pdftex]{hyperref,graphicx} 
\usepackage{cite}
\usepackage[subrefformat=parens]{subcaption}
\usepackage{comment}
\usepackage{fancyhdr} 
\usepackage{amssymb}

\hypersetup{
	colorlinks=true,
	linkcolor=blue,
	bookmarks=true,
	bookmarksnumbered=true,
	pdfborder={0 0 0},
	bookmarkstype=toc
}

\newcommand{\ii}{{\rm i}}
\newcommand{\f}[1]{f_{\rm #1}}
\newcommand{\Om}[1]{\Omega_{\rm #1}} 
\newcommand{\om}[1]{\omega_{\rm #1}} 
\newcommand{\rh}[1]{\rho_{\rm #1}}
\newcommand{\G}{\Gamma}

\renewcommand{\Eref}[1]{Eq.~(\ref{#1})}
\renewcommand{\Fref}[1]{Fig.~\ref{#1}}

\begin{document}
\title[Design and experimental demonstration of a laser modulation system for future gravitational-wave detectors]{Design and experimental demonstration of a laser modulation system for future gravitational-wave detectors}
\author{Kohei Yamamoto$^1$\footnote{Present address: Max Planck Institute for Gravitational Physics (Albert Einstein Institute) and Institute for Gravitational Physics, Leibniz Universit\"{a}t Hannover, Callinstr. 38, 30167 Hannover, Germany}, Keiko Kokeyama$^2$, Yuta Michimura$^3$, Yutaro Enomoto$^3$, Masayuki Nakano$^4$, Gui-Guo Ge$^5$, Tomoyuki Uehara$^{6, 7}$, Kentaro Somiya$^8$, Kiwamu Izumi$^9$, Osamu Miyakawa$^2$, Takahiro Yamamoto$^2$, Takaaki Yokozawa$^2$, Yuta Fujikawa$^{10}$, Nobuyuki Fujii$^{10}$, Takaaki Kajita$^1$}
\address{$^1$ Institute for Cosmic Ray Research (ICRR), Research Center for Cosmic Neutrinos (RCCN), The University of Tokyo, Kashiwa City, Chiba 277-8582, Japan.}
\address{$^2$ Institute for Cosmic Ray Research (ICRR), KAGRA Observatory, The University of Tokyo, Kamioka-cho, Hida City, Gifu 506-1205, Japan.}
\address{$^3$ Department of Physics, The University of Tokyo, Bunkyo-ku, Tokyo 113-0033, Japan}
\address{$^4$ Department of Physics, University of Toyama, Toyama City, Toyama 930-8555, Japan}
\address{$^5$ State Key Laboratory of Magnetic Resonance and Atomic and Molecular Physics, Wuhan Institute of Physics and Mathematics, Chinese Academy of Sciences, West No. 30, Xiaohongshan, Wuhan, 430071, China}
\address{$^6$ Department of Communications, National Defense Academy of Japan, Yokosuka City, Kanagawa 239-8686, Japan}
\address{$^7$ Department of Physics, University of Florida, Gainesville, FL 32611, United States of America}
\address{$^8$ Graduate School of Science and Technology, Tokyo Institute of Technology, Meguro-ku, Tokyo 152-8551, Japan}
\address{$^9$ JAXA Institute of Space and Astronautical Science, Chuo-ku, Sagamihara CIty, Kanagawa 252-0222, Japan}
\address{$^{10}$ Faculty of Engineering, Niigata University, Nishi-ku, Niigata City, Niigata 950-2181, Japan}
\ead{kohei.yamamoto@aei.mpg.de}
\begin{abstract}
Detuning the signal-recycling cavity length from a cavity resonance significantly improves the quantum noise beyond the standard quantum limit, while there is no km-scale gravitational-wave detector successfully implemented the technique. The detuning technique is known to introduce great excess noise, and such noise can be reduced by a laser modulation system with two Mach-Zehnder interferometers in series. This modulation system, termed Mach-Zehnder Modulator (MZM), also makes the control of the gravitational-wave detector more robust by introducing the third modulation field which is non-resonant in any part of the main interferometer. On the other hand, mirror displacements of the Mach-Zehnder interferometers arise a new kind of noise source coupled to the gravitational-wave signal port. In this paper, the displacement noise requirement of the MZM is derived, and also results of our proof-of-principle experiment is reported.\par

\vspace{\baselineskip}
\noindent{\it Keywords: gravitational waves, laser interferometer, modulation, detuning}
\end{abstract}
\pagestyle{fancy}

\section{Introduction}\label{sec:intro}
The gravitational-wave (GW) events from binary black holes and binary neutron stars have been detected by the LIGO-Virgo collaboration~\cite{ref:gw150914,ref:gw151226,ref:gw150914new,ref:gwO1,ref:gw170104,ref:gw170814,ref:gw170608,ref:gw170817},
marking the dawn of the GW astronomy. These second-generation ground-based GW detectors~\cite{ref:ligo,ref:virgo,ref:kagra} deploy
so-called signal recycling cavity~\cite{ref:sr1,ref:sr2,ref:sr3} (SRC)
at the anti-symmetric (AS) port, which is the GW signal port, as shown in \Fref{fig:mainifo}.
By tuning the resonance condition of the SRC,
the frequency response of the detector for incoming GW signals can be tuned.
This way, a science outcome for particular astrophysical GW events can be maximized.

The laser interferometer is a complex system with multiple optical cavities, 
in addition to the km scale Fabry-Perot cavities.
To operate the laser interferometer as a GW detector,
all the length degrees of freedom of these cavities must be actively feedback controlled.
Error signals for the control are obtained
by beating between the carrier light field and phase-modulated (PM) sideband field,
based on the Pound-Drever-Hall-type readout scheme~\cite{ref:pdh}.
Detuning of the SRC is also done by tuning the position of the signal recycling mirror
through the control system.
Thus, the laser modulation system is indispensable for the current and future GW detectors.

Advanced LIGO~\cite{ref:ligo} currently operates the detectors at a particular signal recycling condition
called broadband resonant sideband extraction (BRSE or simply RSE)~\cite{ref:brse1,ref:brse2,ref:brse3}
for a broader observation bandwidth. While KAGRA~\cite{ref:kagra} also plans to run with BRSE,
also the other signal-recycling cavity condition, so called Detuned RSE (DRSE), is proposed.
A major advantage of DRSE is to surpass the quantum noise
below the standard quantum limit~\cite{ref:drse1,ref:drse2}.
On the other hand, a previous study~\cite{ref:ueda} pointed out that the DRSE configuration
enhances two kinds of unwanted noise couplings, {\it i.e.} sensing noise at a photodetector (PDN)
and oscillator phase noise (OPN), resulting in contaminations to the GW sensitivity.
The cause of these couplings is due to a conversion of the PM sideband
to an amplitude modulation (AM) sideband in the SRC.
If it is unaddressed, the oscillator phase stability has to be as good as
-180\,dBc at 100\,Hz~\cite{ref:ueda}, which is practically unreachable.\par

One approach to relax this requirement is to add an AM sideband on the input light field
so that the undesired AM components arisen in the SRC~\cite{ref:ueda} are cancelled.
In general, there are a few options to generate AM sidebands:
One is to utilize birefringence of crystals of electro-optic modulators (EOM)~\cite{ref:amgen2}.
However, this method is greatly affected by temperature fluctuations
and not suitable for GW detectors. An electro-absorption modulator is also a candidate,
which has been studied in telecommunications.
To implement to GW detectors, further R\&Ds are necessary.
For example, applications on high power lasers have to be developed.
In either case, the conventional amplitude modulation comes along with the loss of the carrier power.
Such power loss is not preferable for GW detectors from the perspective of shot noise.
In this paper, we propose a laser modulation system using Mach-Zehnder interferometers (MZIs),
termed Mach-Zehnder Modulator (MZM).
Generally, MZM has been studied in telecommunication because of its versatility~\cite{ref:amgen3}.
Our MZM system is designed to generate AM sidebands while avoiding the laser power loss
by introducing a large asymmetry between the two arms in one of the two MZIs.
The detail will be explained in \sref{sub:concept}.
In addition to the AM sidebands for the cancellation of the undesired AM components caused by the DRSE,
our MZM system also provides another AM sideband field which will not be resonant
in any part of the main interferometer. The non-resonant sideband field will give us
stable error signals for particular length degrees of freedom (DoF) during lock acquisitions.
A technique similar to the non-resonant sideband had been employed in initial LIGO~\cite{ref:nonreso}.\par

There is a technical challenge in the MZM system as a laser modulation system in GW detectors.
Displacements of optics in the MZM system introduce a new noise couplings to the GW readout channel. This paper presents the optical design of our MZM system
together with the displacement noise requirement of the MZM system
which does not degrade the astrophysical reach of the GW detectors.
Our MZM system has an ability to control the relative magnitude and phase
between the PM and newly imposed AM components.
We conducted the proof-of-principle experiment to confirm this functionality.
Also, the current displacement noise level has been evaluated.
Based on the result, we investigate the future mitigation options.

\section{Laser modulation system}\label{sec:mzm}
\subsection{Motivation}\label{sub:motiv}
\begin{figure}[h]
	\centering
	\includegraphics[width=12cm]{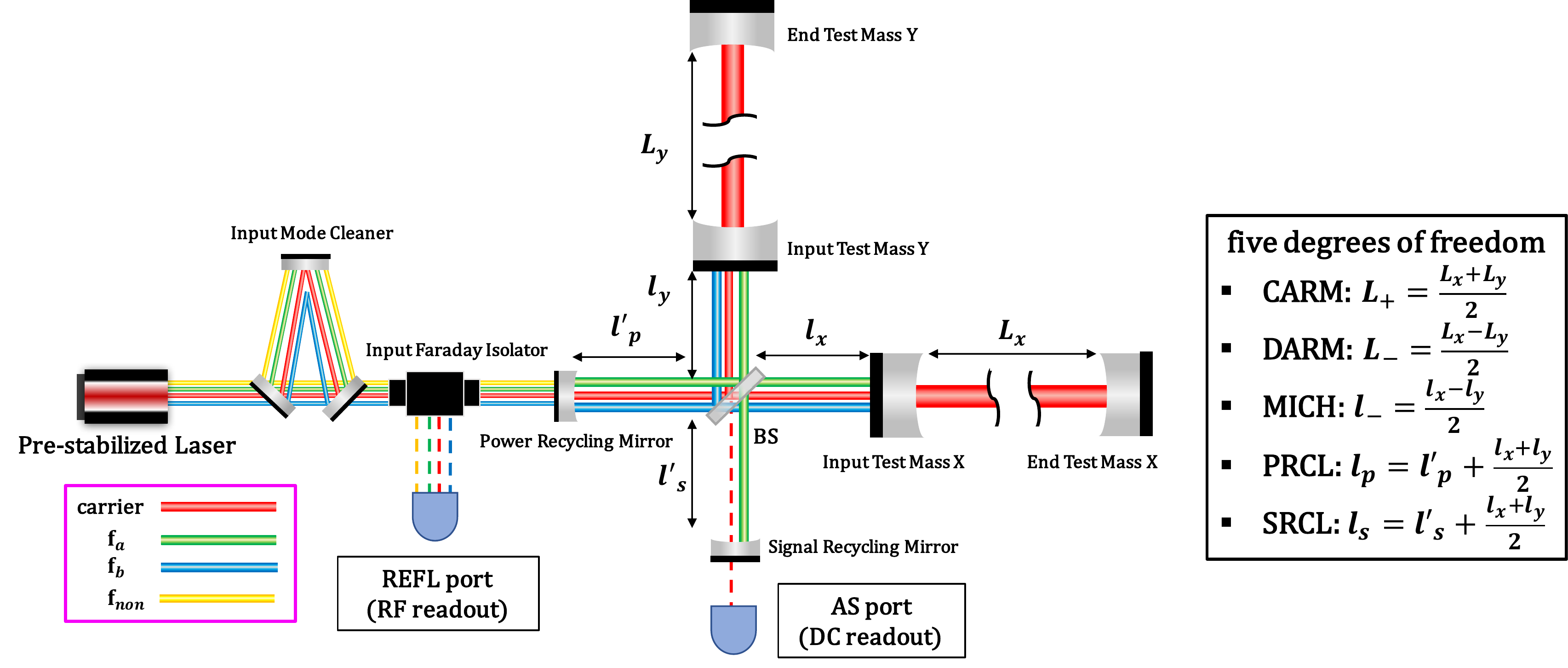}
	\renewcommand{\baselinestretch}{1.0}
	\caption{Schematic of a second-generation ground-based GW detector. In addition to the sidebands at $\f{a}$ and $\f{b}$, the figure also includes the non-resonant sideband, $\f{non}$. On the right panel, the five length degrees of freedom are defined; CARM: the common arm cavity length. DARM: the differential arm cavity length (GW channel). MICH: the differential length of the short Michelson interferometer part. PRCL: the power recycling cavity length. SRCL: the signal recycling cavity length. Gravitational-wave signals are detected by a DC coupled PD (DCPD) at the AS port (DC readout).}
	\label{fig:mainifo}
\end{figure}

The second-generation ground-based GW interferometers have five length DoFs
to be controlled on proper resonant conditions, as shown in \Fref{fig:mainifo}.
Feedback control is applied to {\it lock} each DoF at the operating point.
Error signals for each control loop should be as linear as possible,
and also the coefficient of the response should be as large as possible.
To obtain such good error signals, the RF modulation and demodulation technique is used
for the four DoFs except for the differential motion of the arm cavities (DARM DoF, the GW channel).
Moreover, the error signals of the five DoFs must be diagonalized well
to avoid any cross couplings between the controlled DoFs.
To address the complexity of the interferometer controls,
multiple modulation frequencies and signal extraction ports are used, as depicted in \Fref{fig:mainifo}.
For the signal extractions, the input laser field is phase modulated at two independent RF frequencies~\cite{ref:twofreq1, ref:twofreq2, ref:twofreq3}.
The modulated input field is written as
\begin{eqnarray}
E_{\rm in}=E_{0}\e^{\ii\omega t}[1+(\G^{\f{a}}_{\rm AM}+\ii \G^{\f{a}}_{\rm PM})\sin\Om{a}t+\ii \G^{\f{b}}_{\rm PM}\sin\Om{b}t+\G^{\f{non}}_{\rm AM}\sin\Om{non}t], \label{eq:input}
\end{eqnarray}
where $\Om{m} ({\rm m=a,b,non})$ is the modulation frequencies
and $\G^{\f{m}}_{\i}$ (i = PM or AM) is the modulation index.
The first term is the carrier field, the second and third terms are AM and PM components
at frequency of $\f{a}$, respectively,
the forth term is a PM component at $\f{b}$, and the fifth term is an AM component at $\f{non}$.
The frequency of $\f{a}$ is chosen so to resonate in the SRC,
to carry information of the SRC length.
This equation only includes the first order sidebands
because the modulation indices are typically small enough to omit the higher order sidebands.\par

This light field delivers two key features. First, the input light is to coherently
add an AM component to the PM component at $\f{a}$.
The complex reflectivity of the main interferometer is a function of modulation frequencies:
\begin{eqnarray}
r(\Om{m};\Phi)=A(\Om{m};\Phi)\exp[\ii\alpha(\Om{m};\Phi)], \label{eq:refle}
\end{eqnarray}
where $\Phi$ is the detune phase in the SRC;
$A$ and $\alpha$ are the magnitude and phase in a complex plane, respectively.
The magnitude $A$ always meets $0\leqq A\leqq 1$ by definition and $\alpha$ is zero on resonance.
Because $\alpha$ has non-zero value when a cavity length deviates from the resonance,
the DRSE generates an undesirable AM component at $\f{a}$, {\it i.e.}
$\mathrm{Re}[\ii A\e^{\ii \alpha}\G^{\f{a}}_{\rm PM}\sin\Om{a}t]$.
This component results in two noise coupling mechanisms,
PDN and OPN at the interferometer reflection (REFL) port.
The newly-introduced AM component at $\f{a}$ in  \Eref{eq:input} solves
this problem if well-tuned to cancel out the unwanted AM component:
\begin{eqnarray}
\mathrm{Re}[A\e^{\ii \alpha}(\G^{\f{a}}_{\rm AM}+\ii \G^{\f{a}}_{\rm PM})]=0.\label{eq:amcancel}
\end{eqnarray}\par
The second key feature is the non-resonant AM sideband at $\f{non}$.
During the lock acquisition phase, a stable reference field which is not affected by any resonant conditions in the main interferometer
is useful to obtain well-decoupled error signals of length DoFs.
The non-resonant AM sideband acts as such reference field.
Particularly for the central part of the main interferometer, error signals can be extracted
by beating between sidebands at $\f{a}$ (or $\f{b}$) and $\f{non}$
(therefore the demodulation frequencies are $\f{non}-\f{a}$ and $\f{non}-\f{b}$),
without signal couplings between the central part and arm cavities in principle.
To beat with the PM sidebands at $\f{a}$ or $\f{b}$,
the sideband field at $\f{non}$ has to be an AM sideband. As mentioned above, the non-resonant sideband at $\f{non}$ will be used only during the lock acquisition phase because the main component of this signal is a beatnote between two sideband fields. Its signal-to-noise ratio will not be as good as the conventional signal, which is a beatnote between the carrier and sideband fields.\par

\subsection{Concept of the MZM system}\label{sub:concept}
\begin{figure}[h]
	\centering
	\includegraphics[width=10cm]{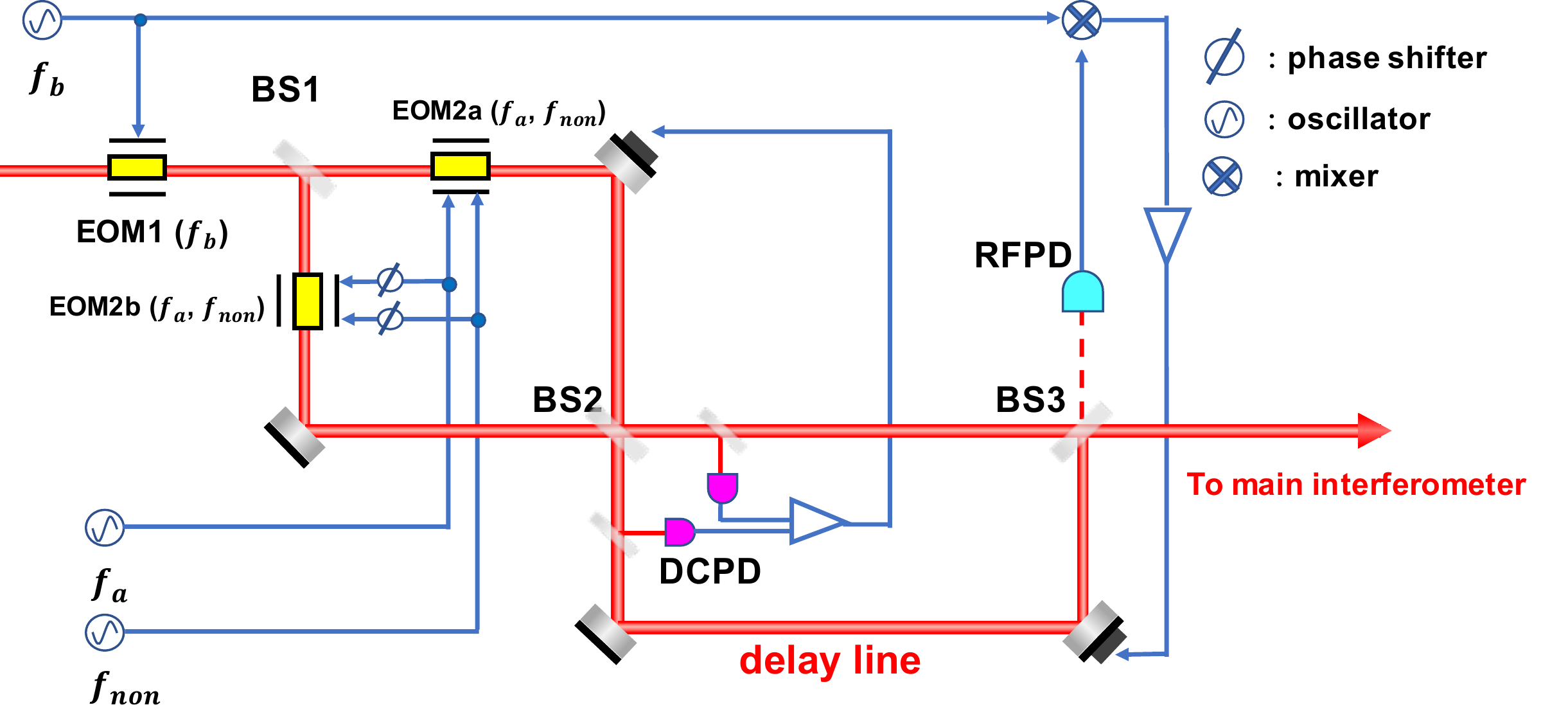}
	\renewcommand{\baselinestretch}{1.0}
	\caption[Schematic of the MZM]{Schematic view of the MZM system. The light and RF signals are shown in red and blue, respectively. The first MZI is locked to the mid-fringe with the differential method using two photo-detectors with DC coupled PDs (DCPDs). The second MZI is locked to the dark-fringe using the sidebands at $\f{b}$. The role of the first MZI is to generate the AM sideband, and the role of the second MZI is to reuse the beam lost at the output of the first MZI.}
	\label{fig:mzmconfig}
\end{figure}

Our MZM system is optimized to generate the input light field written in \Eref{eq:input}.
\Fref{fig:mzmconfig} shows the schematic view of the concept.
After the first EOM (EOM1) modulating the light in phase at the frequency of $\f{b}$
and at the frequency for the input mode cleaner control, the light passes through two MZIs placed in series.
EOM2a and EOM2b driven by a common function generator are to apply
$\f{a}$ and $\f{non}$,
and are symmetrically placed in each path in the first symmetrical MZI.
Each EOM in the path is high-power-applicable
and has two pairs of electrodes to apply two independent modulation frequencies.

The first MZI is locked to the mid-fringe to generate the PM and AM sidebands
and the second (asymmetrical) MZI is locked to the dark-fringe
so that the half of the light power lost at the output of the first MZI can be reused.
All the modulation fields can be summarized in one formula: 
\begin{eqnarray}
E_{\rm out}=E_{0}\e^{\ii \omega t} \left[1+\G\sin\frac{\phi}{2}\sin\frac{\theta}{2}\cos\left(\Om{m}t+\frac{\theta+\phi-\pi}{2}\right)\right.\nonumber\\
\left.\hspace{29mm}+\ii \G\cos\frac{\phi}{2}\cos\frac{\theta}{2}\cos\left(\Om{m}t+\frac{\theta+\phi-\pi}{2}\right)\right],
\label{eq:mzmoutput}
\end{eqnarray} 
where the first term is the carrier field, the second and third terms are the AM and PM components
at the certain modulation frequencies, respectively.
The modulation terms include two inner parameters: $\phi$ is a phase difference
between EOM2a and EOM2b in the first MZI. $\theta$ is a phase delay introduced
by the length asymmetry in the second MZI.
\Eref{eq:mzmoutput} suggests that asymmetry in the second MZI is intrinsically important for the AM-generating modulator with the power-lossless feature.
If there is no asymmetry ($\theta = 0$), the second term in \Eref{eq:mzmoutput}, {\it e.g.} the AM component, vanishes.
Note that $\phi$ can be changed by a phase shifter, while $\theta$ is fixed parameter depending on the delay line length.

\begin{figure}[h]
	\begin{minipage}[h]{0.52\linewidth}
		\centering
		\includegraphics[width=7.0cm]{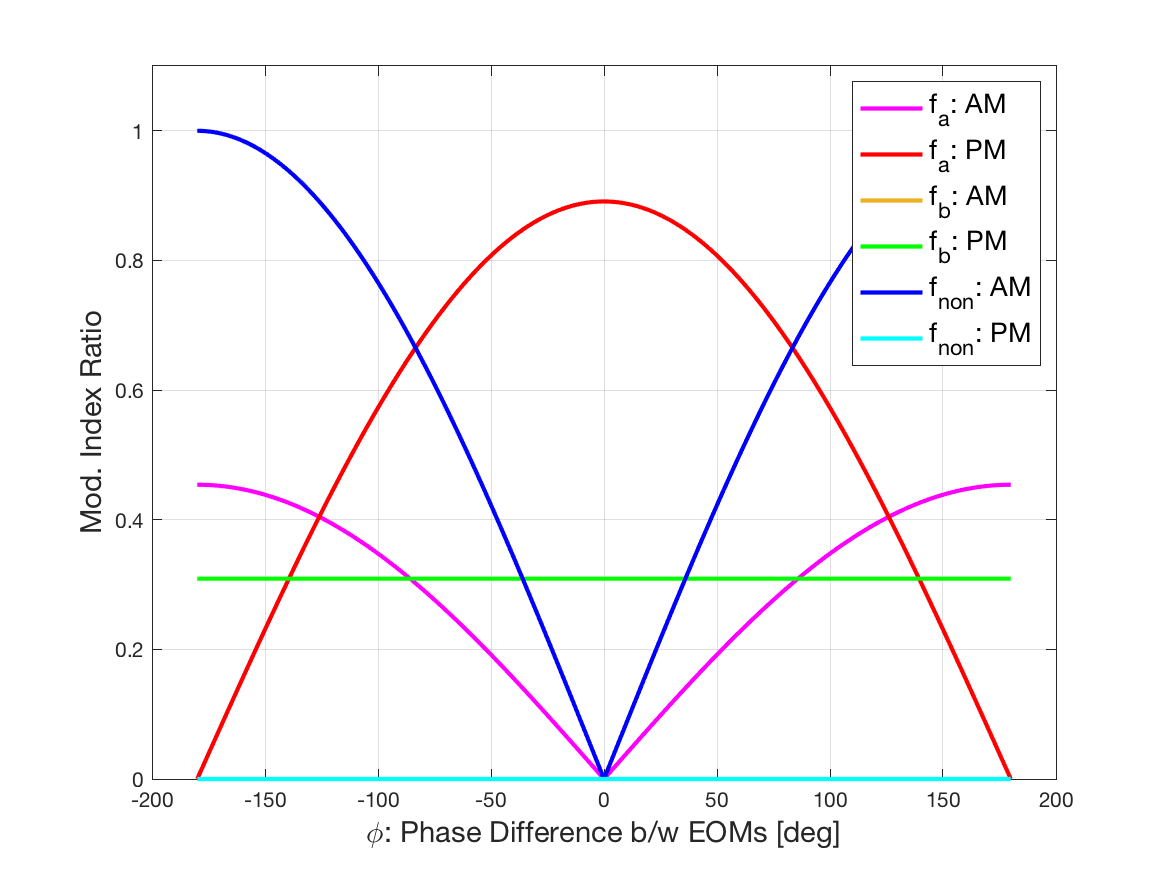}
		\subcaption{Bright port}
		\label{fig:indexphi_a}
	\end{minipage}
	\begin{minipage}[h]{0.52\linewidth}
		\centering
		\includegraphics[width=7.0cm]{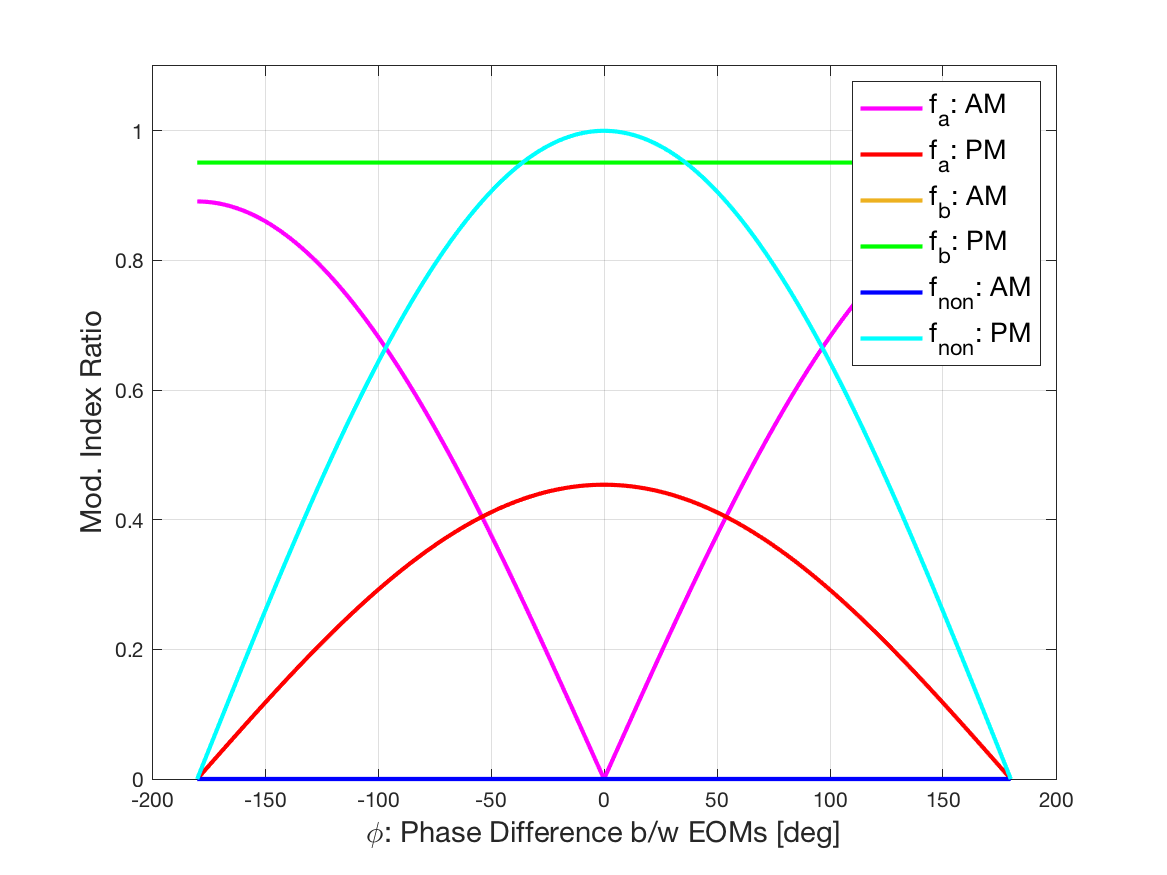}
		\subcaption{Dark port}
		\label{fig:indexphi_b}
	\end{minipage}
	\renewcommand{\baselinestretch}{1.0}
	\caption{Modulation index ratio between AM and PM sidebands over the phase difference $\phi$:
	Magenta trace is the AM component at $\f{a}$, red trace is the PM component at $\f{a}$,
	yellow trace is the AM component at $\f{b}$, green trace is the PM component at $\f{b}$,
	blue trace is the AM component at $\f{non}$, and light blue trace is the PM component at $\f{non}$.
	The AM component at $\f{b}$ and PM component at $\f{non}$ are zero regardless of $\phi$.
	By increasing $\phi$ from zero, we can produce the AM component at $\f{a}$ and $\f{non}$.
	90\,\% of the power of the modulation at $\f{b}$ goes to the dark port.}
	\label{fig:indexphi}
\end{figure}

A remarkable advantage of the MZM system is the various options of modulations achieved by these parameters:
If the length of the delay line is chosen so that $\theta=\pi$ for $\f{non}$,
the MZM system can prohibit a PM component at $\f{non}$ by the configuration.
Moreover, $\phi$ enables to control the relative magnitude between the PM and AM components for the sideband at $\f{a}$,
and the amplitude of the AM at $\f{non}$ by a phase shifter anytime as shown in \Fref{fig:indexphi}.

Furthermore, it is possible to introduce an additional parameter in the MZM system, {\it i.e.} the asymmetry of the amount of the modulations produced by the EOMs in the first MZI. This is necessary if the reflectivity and transmissivity of the BS in the main interferometer are not equal, as pointed out in~\cite{ref:ueda}. In this case, \Eref{eq:mzmoutput} can be written as
\begin{eqnarray}
E_{\rm out}=E_{0}{\rm e}^{\ii\omega t}\left[1+\frac{1}{2}\sqrt{2\G(\G+\alpha)(1-\cos\phi)+\alpha^{2}}\sin\frac{\theta}{2}\cos\left(\Om{m}t+\frac{\theta}{2}+\rh{AM}\right)\right.\nonumber\\
\hspace{29mm}\left.+\frac{1}{2}\ii\sqrt{2\G(\G+\alpha)(1+\cos\phi)+\alpha^{2}}\cos\frac{\theta}{2}\cos\left(\Om{m}t+\frac{\theta}{2}+\rh{PM}\right)\right],\nonumber\\
\rh{AM}=\arctan\left(\frac{-\G\sin\phi}{\G(1-\cos\phi)+\alpha}\right), \hspace{5mm}\rh{PM}=\arctan\left(\frac{-\G(1+\cos\phi)+\alpha}{\G\sin\phi}\right),
\label{eq:mzmoutasym}
\end{eqnarray}
where $\alpha$ is the asymmetry of the modulation index produced by EOMs. Note that when $\alpha = 0$, \Eref{eq:mzmoutasym} agrees with \Eref{eq:mzmoutput}.

\subsection{Technical challenge: coupling to the AS port}\label{sub:drawback}

As mentioned in \Sref{sub:motiv}, the purpose of generating the AM sideband by the MZM is to cancel the unwanted AM component at the REFL port. This is expected to greatly alleviate the amount of SRCL and/or MICH noise contaminating the DARM DoF (which is the GW DoF) through inherent couplings of these control loops.

On the other hand, the MZM practically generate the displacement noise of the optics in MZIs, which goes into the main interferometer as an amplitude and phase noise of the sidebands. They introduce a new noise coupling to the GW channel; from the MZI displacement to the fluctuation of the DC power at the AS port. In the following section, we will discuss the new noise coupling.

\section{Displacement noise}\label{sec:dispnoise}
Using Optickle\cite{ref:optickle}, the interferometer simulation implemented in MATLAB,
the requirements on displacement noise were derived.
The numerical simulation was performed with the KAGRA optical parameters
summarized in \Tref{tab:simuparamifo} and \ref{tab:simuparammod} in the appendix. 

\subsection{Characteristic}\label{sub:analytic}
Before deriving requirements, it is useful to discuss about the conversions from displacements
of the MZM optics into the sideband noise in this new modulation system.

\begin{figure}[h]
	\centering
	\includegraphics[width=9cm]{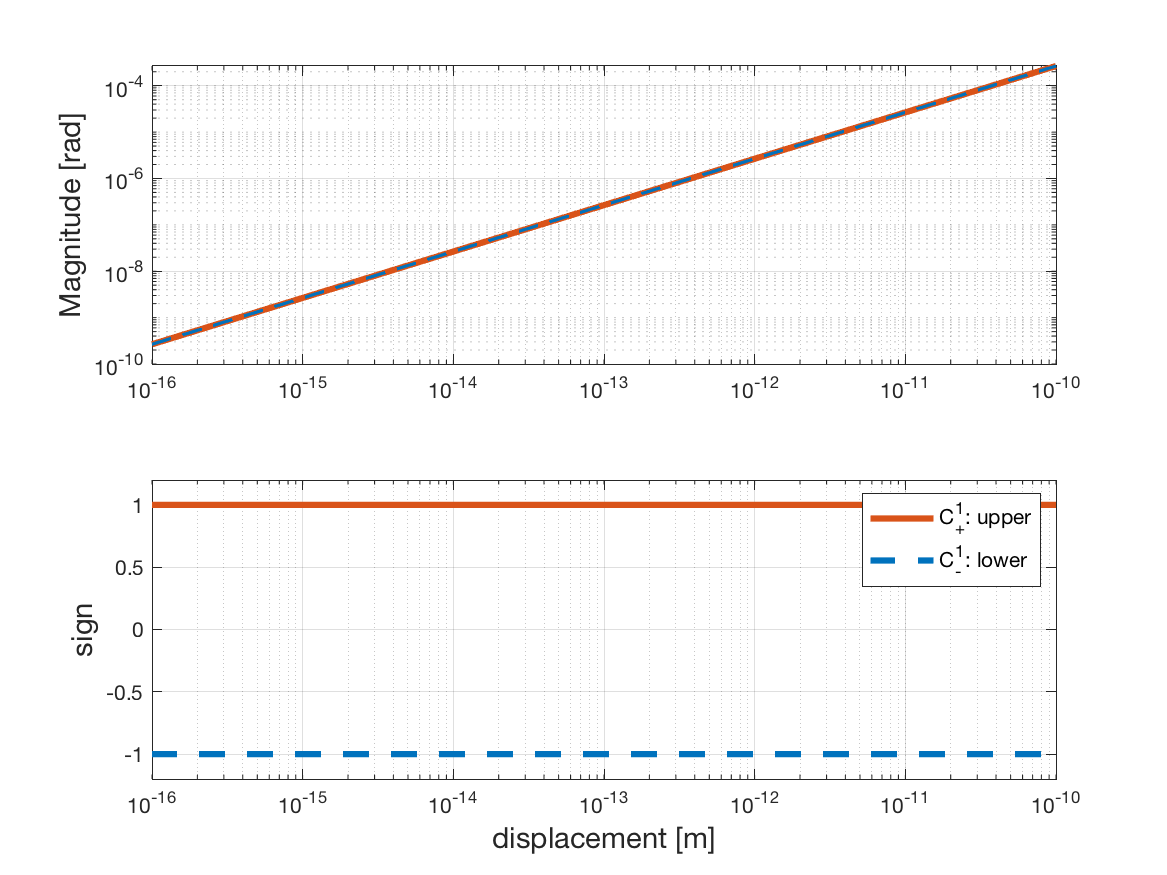}
	\renewcommand{\baselinestretch}{1.0}
	\caption{Phase noises of the upper and lower sidebands caused by the displacement in the first MZI: they have almost the equal magnitudes with a very high precision and the opposite signs. The amplitude noise and those in the second MZI also have the same characteristic. Here the $\f{a}$ frequency and the phase is 16.88\,MHz and $\phi = 100$\,deg, respectively.}
	\label{fig:sdpnvsdisp}
\end{figure}

Modulated light can be written in an abstract form as,
\begin{eqnarray}
E&=E_{0}{\rm e}^{i\omega t}\left[1+C_{+}{\rm e}^{i\Om{m}t}+C_{-}{\rm e}^{-i\Om{m}t}\right],\label{eq:pmamconversion}
\end{eqnarray}
where $C_{+}$ ($C_{-}$) is a complex amplitude of the upper (lower) sideband.
Once we have the displacement noise, $\delta \ell$,
in the first or second MZI, $C_{+}$ and $C_{-}$ are concretely derived as the functions of $\delta \ell$:
\begin{eqnarray}
C^{1}_{\pm}(\delta \ell;\pm\Om{m},\pm\phi,\pm\theta)\nonumber\\
\hspace{2mm}=\frac{\G\e^{i\left(\frac{\theta}{2}+\frac{\Om{m}\delta \ell}{2c}\right)}}{2\sqrt{2}\cos\left(\frac{\omega\delta \ell}{2c}\right)}\left[\e^{i\frac{\om{+}\delta \ell}{2c}}\cos\left(\frac{\theta}{2}-\frac{\pi}{4}\right)+\e^{i\left(\phi-\frac{\om{+}\delta \ell}{2c}\right)}\cos\left(\frac{\theta}{2}+\frac{\pi}{4}\right)\right],\label{eq:c1pm}\\
C^{2}_{\pm}(\delta \ell;\pm\Om{m},\pm\phi,\pm\theta)\nonumber\\
\hspace{2mm}=\frac{\G\e^{i\left(\frac{\theta}{2}+\frac{\Om{m}\delta \ell}{2c}\right)}}{2\sqrt{2}\cos\left(\frac{\omega\delta \ell}{2c}\right)}\left[\cos\left(\frac{\theta}{2}+\frac{\om{+}\delta \ell}{2c}-\frac{\pi}{4}\right)+\e^{i\phi}\cos\left(\frac{\theta}{2}+\frac{\om{+}\delta \ell}{2c}+\frac{\pi}{4}\right)\right],\label{eq:c2pm}
\end{eqnarray}
where $\om{\pm}=\omega\pm\Om{m}$.
The superscripts on $C$ denote the first or second MZI, and $+ (-)$ denotes an upper (lower) sideband.
Note that when $\delta \ell = 0$, Eqs. (\ref{eq:c1pm}) and (\ref{eq:c2pm}) derive the fields of \Eref{eq:mzmoutput}.
Eqs. (\ref{eq:c1pm}) and (\ref{eq:c2pm}) shows that the noise contributions of the displacement fluctuation $\delta \ell$
in the upper and lower sidebands have almost the equal magnitudes but the opposite signs
within a reasonable range of $\delta \ell$, {\it e.g.} $10^{-16}-10^{-10}$\,m
as shown in \Fref{fig:sdpnvsdisp}\footnote{We assume the size of the modulation system is small enough to neglect the frequency-dependence of the noise coupling.}.\par

As for a function generator, the phase (amplitude) noise, $\delta \phi$ ($\delta \G$), can be simply introduced
by the replacement in \Eref{eq:mzmoutput}; $\Om{m}t \rightarrow \Om{m}t+\delta\phi$ ($\G \rightarrow \G+\delta \G$).
This is because we use a single function generator for the two EOMs in the first MZI.
This results in the fact that the phase (amplitude) noise of a function generator driving EOMs
still keeps the original feature for the output light of the MZM system,
{\it i.e.} identical magnitudes but opposite (equal) signs between the upper and lower sidebands.\par

\begin{table}
	\caption{\label{tab:mzmcoup}Characteristic of sideband noise at the output of the MZM from different noise sources; function generator, first and second MZI displacements. The number in the row ``diff. in magnitude'' is the relative magnitude differences between the upper and lower sidebands at the phase difference between the EOMs of 100\,deg.}
	\begin{indented}
		\item[]\begin{tabular}{@{}lclclclclclclcl}
			\br
			& \multicolumn{3}{c}{Phase Noise}&\multicolumn{3}{c}{Amplitude Noise}\\
			\mr
			factor&SG&first disp.&second disp.&SG&first disp.&second disp.\\
			\mr
			diff. in magnitude &0&$2.6\times10^{-7}$&$5.6\times10^{-8}$&0&$5.0\times10^{-7}$&$2.4\times10^{-6}$ \\
			opposite sign &$\bigcirc$&$\bigcirc$&$\bigcirc$&$\times$&$\bigcirc$&$\bigcirc$\\ 
			\br
		\end{tabular}
	\end{indented}
\end{table}

\subsection{Simulation}\label{sub:simulation}
\begin{figure}[h]
	\centering
	\includegraphics[width=8cm]{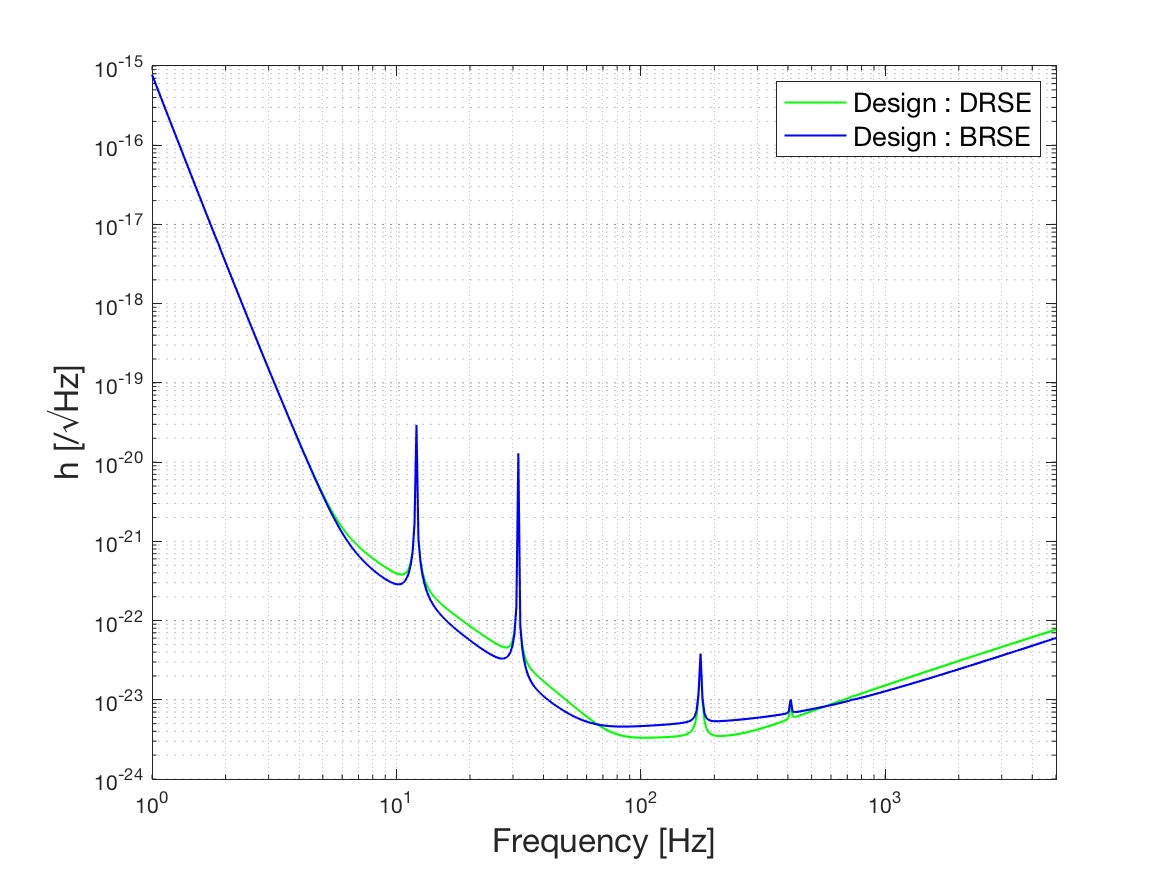}
	\renewcommand{\baselinestretch}{1.0}
	\caption[Sensitivity curves used for the simulation]{Sensitivity curves used for the simulation. All the parameters are summarized in \Tref{tab:simuparamifo} and \Tref{tab:simuparammod}.}
	\label{fig:simusenstivity}
\end{figure}

The requirement on the displacement fluctuation of the MZM system in the frequency domain, $\delta \ell(f)$ is written as,\par
\begin{eqnarray}
\delta \ell(f) = 0.1\times\frac{1}{T_{\rm MZM}(f)}\times T_{\rm DARM}(f)\times h(f)\times L,\label{eq:req}
\end{eqnarray}
where $L$ is the arm length of the main interferometer,
$h(f)$  is the strain sensitivity, $T_{\rm DARM}(f)$ is
the transfer function from the DARM (GW) DoF to the power of the DC coupled PD  (DCPD) at the AS port (GW detection port),
$T_{\rm MZM}(f)$ is the transfer function from the MZM displacement to the AS DCPD power
and 0.1 is a safety factor.
Strain sensitivities of BRSE and DRSE used in the calculation is shown in \Fref{fig:simusenstivity}.

\begin{figure}[h]
	\begin{minipage}[h]{0.52\linewidth}
		\centering
		\includegraphics[width=8.0cm]{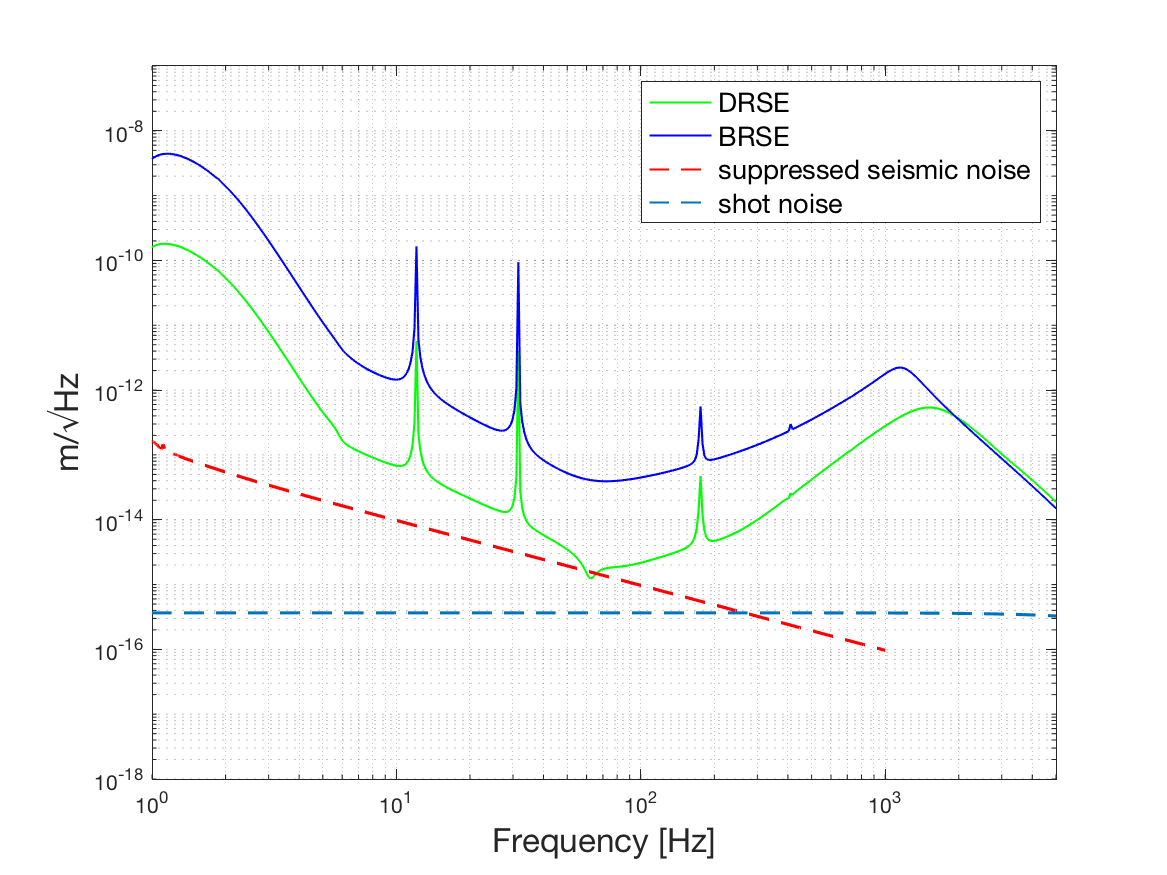}
		\subcaption{first MZI}
		\label{fig:simudispreq_a}
	\end{minipage}
	\begin{minipage}[h]{0.52\linewidth}
		\centering
		\includegraphics[width=8.0cm]{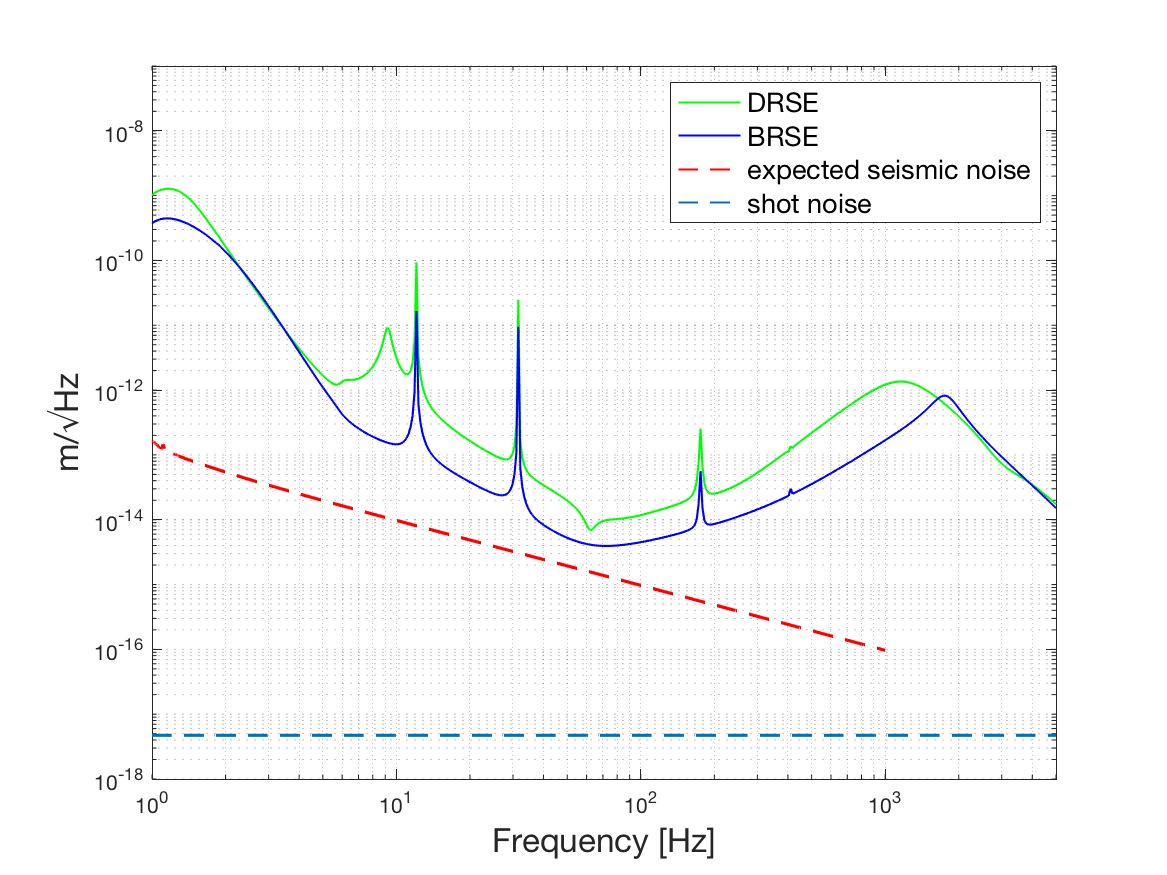}
		\subcaption{second MZI}
		\label{fig:simudispreq_b}
	\end{minipage}
	\renewcommand{\baselinestretch}{1.0}
	\caption{Requirements on the MZM displacement noise.
	For the MZI control, a simple low-pass filter with a unity gain frequency of 10\,kHz for seismic noise suppression,
	and 10\,mW pick-off to the DCPD for the control of the first MZI were assumed.}
	\label{fig:simudispreq}
\end{figure}

\Fref{fig:simudispreq} shows the derived requirements on the displacement noise of both the first and second MZIs
for BRSE and DRSE configurations.
This result confirms the feasibility of the MZM system:
The detection limit set by shot noise is smaller than the displacement noise requirement.\par

\section{Experiment}\label{sec:experiment}
A series of experiments was conducted in the KAGRA site. 
Our first goal was to demonstrate the key functionality of the phase variability shown in \Fref{fig:indexphi},
and the second goal was to evaluate the level of the displacement noise of the MZM.

\subsection{Experimental setup}\label{sub:setup}
\begin{figure}[h]
	\centering
	\includegraphics[width=10cm]{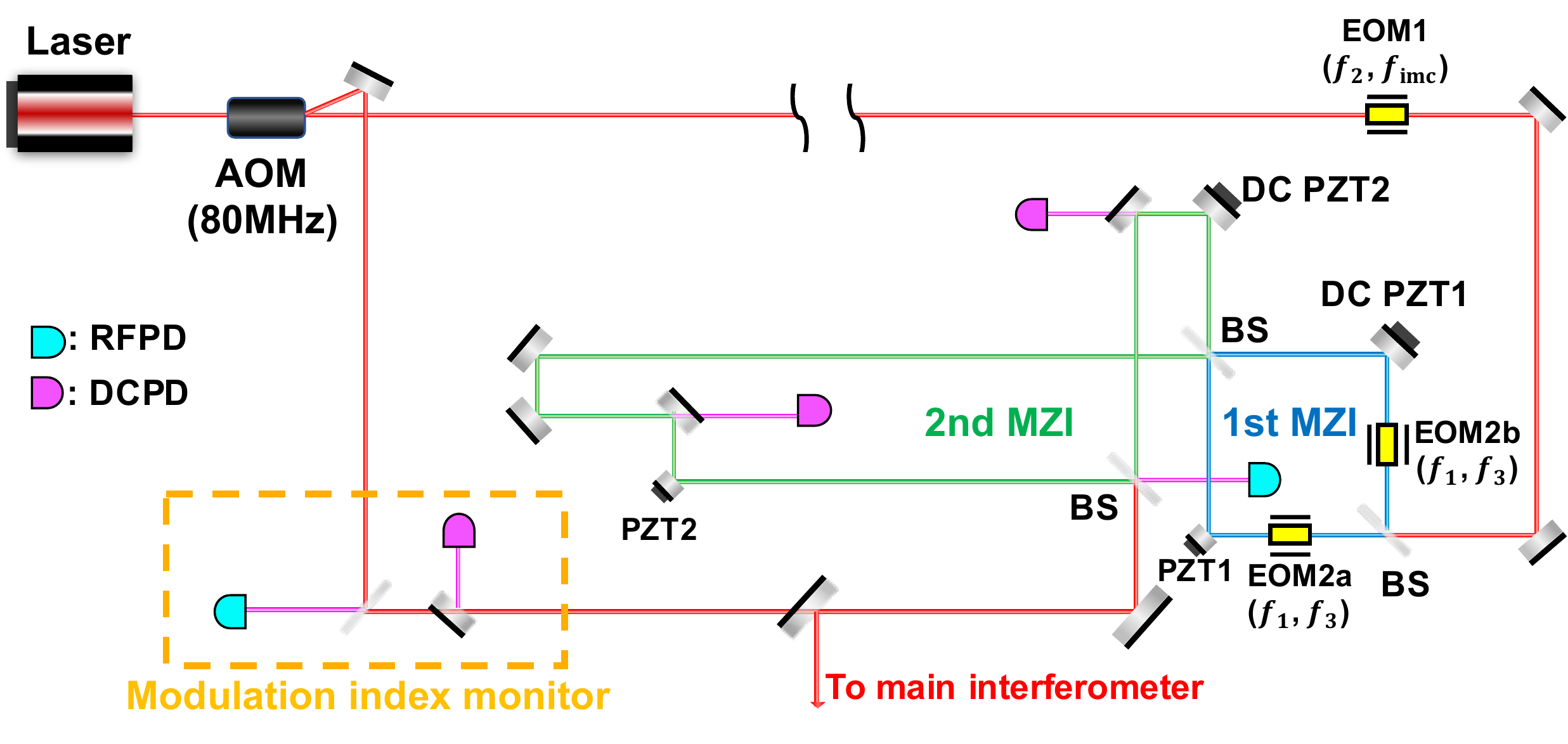}
	\renewcommand{\baselinestretch}{1.0}
	\caption{Schematic of the modulation index measurement. Red: laser on single paths,
	Blue: laser on the first MZI, Green: laser on the second MZI,
	Megenta: Signal laser. The difference in the length
	between two paths on the second MZI was almost 2.66\,m. Laser power was 1.5\,W.}
	\label{fig:setup}
\end{figure}

The setup of the MZM system shown in \Fref{fig:setup} was configured
on the optical table for the laser pre-stabilization systems in KAGRA.
On the delay-line of the second MZI, two lenses were placed for the mode-matching.
As discussed in \sref{sub:concept}, the length of the delay-line was optimized for a pure AM at $\f{3}$:
\begin{eqnarray}
\hspace{20mm}k_{3}l =\theta_{3}=\pi\hspace{3mm}\therefore l \sim 2.66{\rm \hspace{1mm}m}.\label{eq:delayline} 
\end{eqnarray}

To measure the modulation indices of the AM and PM components, we used the beatnote
between the MZM output light and a first diffraction beam of an acousto-optic modulator (AOM) implemented upstream.
Here the AM sideband frequency was 80\,MHz shifted from the carrier light.
As shown in the left bottom in \Fref{fig:setup}, the 80\,MHz-shifted beam
and MZM output beam were mixed at a beamsplitter in the modulation index monitor part
of the setup, and the beatnote was detected by a radio-frequency photodetector (RFPD).
The detected signals were measured by a spectrum analyzer,
and the signal power at the corresponding frequency ({\it e.g.} 80\,MHz - $\f{1}$ for the $\f{1}$ PM sideband,
and $\f{1}$ Hz for the $\f{1}$ AM sideband)
were measured. From the measured signal power,
the modulation index for each component was calibrated.

As for the MZI control, both first and second MZI were controlled by two loops each.
This was to obtain a large control range with good stability during a long-term operation.
The actuators were piezoelectric transducers (PZT) for the control.
The modulation parameters for KAGRA are listed in \Tref{tab:kagramod}.

\begin{table}
	\caption{\label{tab:kagramod}Modulation parameters for KAGRA} 
	\begin{indented}
		\item[]\begin{tabular}{@{}lclclclcl}
			\br
			Name & type & frequency [MHz] & modulation index at PRM\\
			\mr
			$\f{1} (=\f{a})$&AM\&PM&16.88&0.15\\
			$\f{2} (=\f{b})$&PM&45.02&0.05\\
			$\f{3} (=\f{non})$&AM&56.27&0.05\\
			\br
		\end{tabular}
	\end{indented}
\end{table}
\par

\subsection{Calibration of measurements}\label{sub:calib}
To obtain the modulation indices from the measured power,
calibrations were done as follows.

The optical field at the RFPD in the modulation index monitor,
$P_{\rm PD}$, is written as the following formula:
\begin{eqnarray}
P_{\rm PD}&=P_{\omega'}+P_{\omega'-\Om{m}}+P_{\omega'+\Om{m}}+(P_{\rm DC}+P_{\Om{m}}+P_{2\Om{m}}),\label{eq:ppd}\\
\end{eqnarray}
with
\begin{eqnarray}
|P_{\omega'}|&= E_{0}E'_{0},\label{eq:PMmonsignal5}\\
|P_{\omega'-\Om{m}}|&=\frac{E_{0}E'_{0}}{2}\sqrt{\G^{2}_{p}+\G^{2}_{a}-2\G_{p}\G_{a}\sin\rh{AP}},\label{eq:PMmonsignal6}\\
|P_{\omega'+\Om{m}}|&=\frac{E_{0}E'_{0}}{2}\sqrt{\G^{2}_{p}+\G^{2}_{a}+2\G_{p}\G_{a}\sin\rh{AP}},\label{eq:PMmonsignal7}
\end{eqnarray}
where $E'_{0}$ is the amplitude of the first diffraction beam, $\omega'$ is the frequency shifted by the AOM
and $\rh{AP}$ is the phase difference between the AM and PM sidebands.
This comes from the EOM asymmetry $\alpha$ according to Eq. (\ref{eq:mzmoutasym}).\par

From Eqs. (\ref{eq:PMmonsignal5}), (\ref{eq:PMmonsignal6}) and (\ref{eq:PMmonsignal7}),
the total modulation index, $\G_{t}$, being independent from $\rh{AP}$, can be defined as:
\begin{eqnarray}
\G_{t} &\equiv \sqrt{2\times(|P_{\omega'-\Om{m}}|^{2}+|P_{\omega'+\Om{m}}|^{2})}/|P_{\omega'}|\nonumber\\
&=\sqrt{\G^{2}_{p}+\G^{2}_{a}}.\label{eq:PMmonsignal8}
\end{eqnarray}
\par
On the other hand, $\G_{a}$ can be measured from $P_{\Om{m}}$ in Eq. (\ref{eq:ppd})
and the DC power of the MZM output light. The latter was monitored independently
by a DCPD in the modulation index monitor part.
In this manner, all the components, $\G_{t}$, $\G_{a}$ and $\G_{p}$, at a certain modulation frequency can be derived.
Moreover, $\rh{AP}$ is also measured based on the following equation:
\begin{eqnarray}
\sin\rh{AP} = \frac{1}{\G_{a}\G_{p}}\frac{|P_{\omega'-\Om{m}}|^{2}-|P_{\omega'+\Om{m}}|^{2}}{|P_{\omega'}|^2}.\label{eq:PMmonsignal9}
\end{eqnarray}\par

\subsection{Measurements}\label{sub:measurement}
\subsubsection{Phase variability}\label{subsub:phase}
To demonstrate the phase variability of the MZM,
the phase difference between the EOMs in the first MZI was swept
by inserting an additional length on the cable from the function generator to one of the two EOMs.
Then $\G_{t}$, $\G_{a}$ and $\G_{p}$ were derived at each phase difference point.
The experimental parameters are summarized in \Tref{tab:indexvsphiparam}.

\begin{figure}[h]
	\begin{minipage}[h]{0.5\linewidth}
		\centering
		\includegraphics[width=7.0cm]{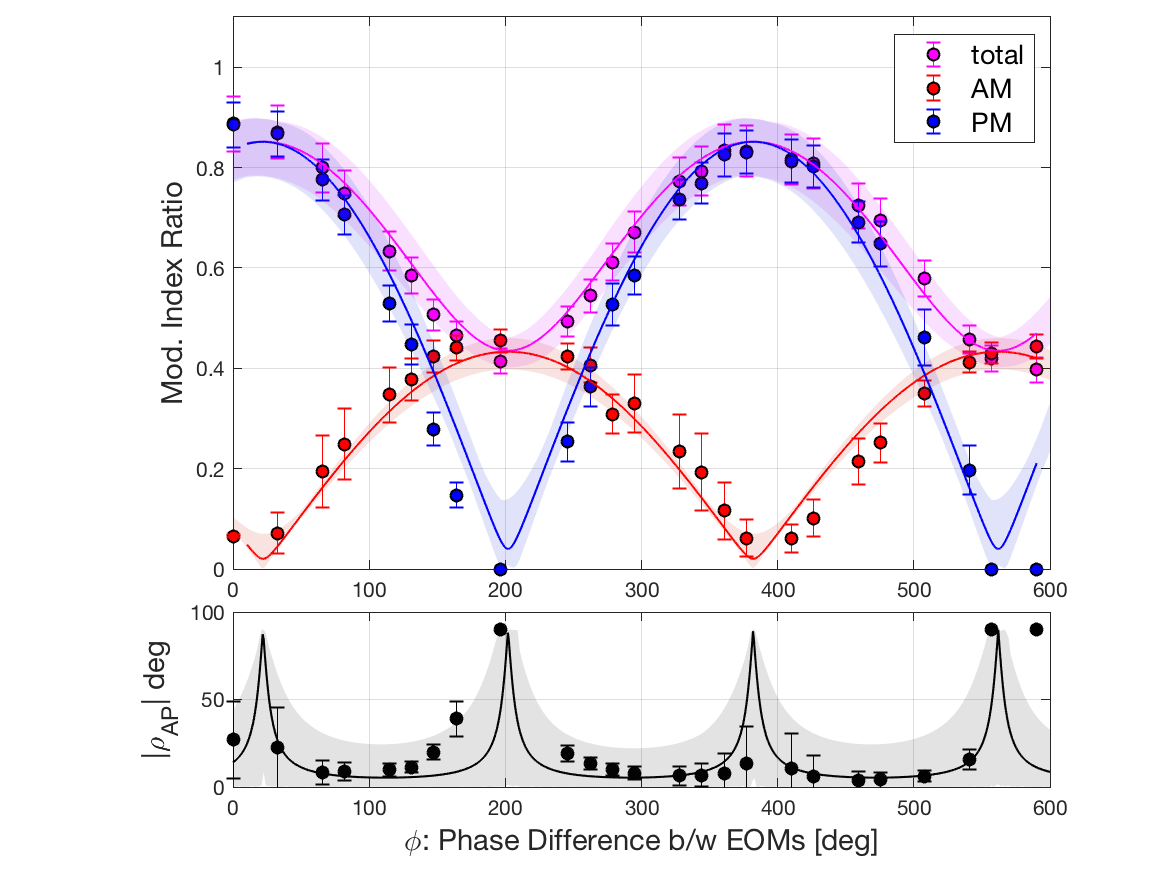}
		\subcaption{$\f{1}$}
		\label{fig:indexvsphi_a}
	\end{minipage}
	\begin{minipage}[h]{0.5\linewidth}
		\centering
		\includegraphics[width=7.0cm]{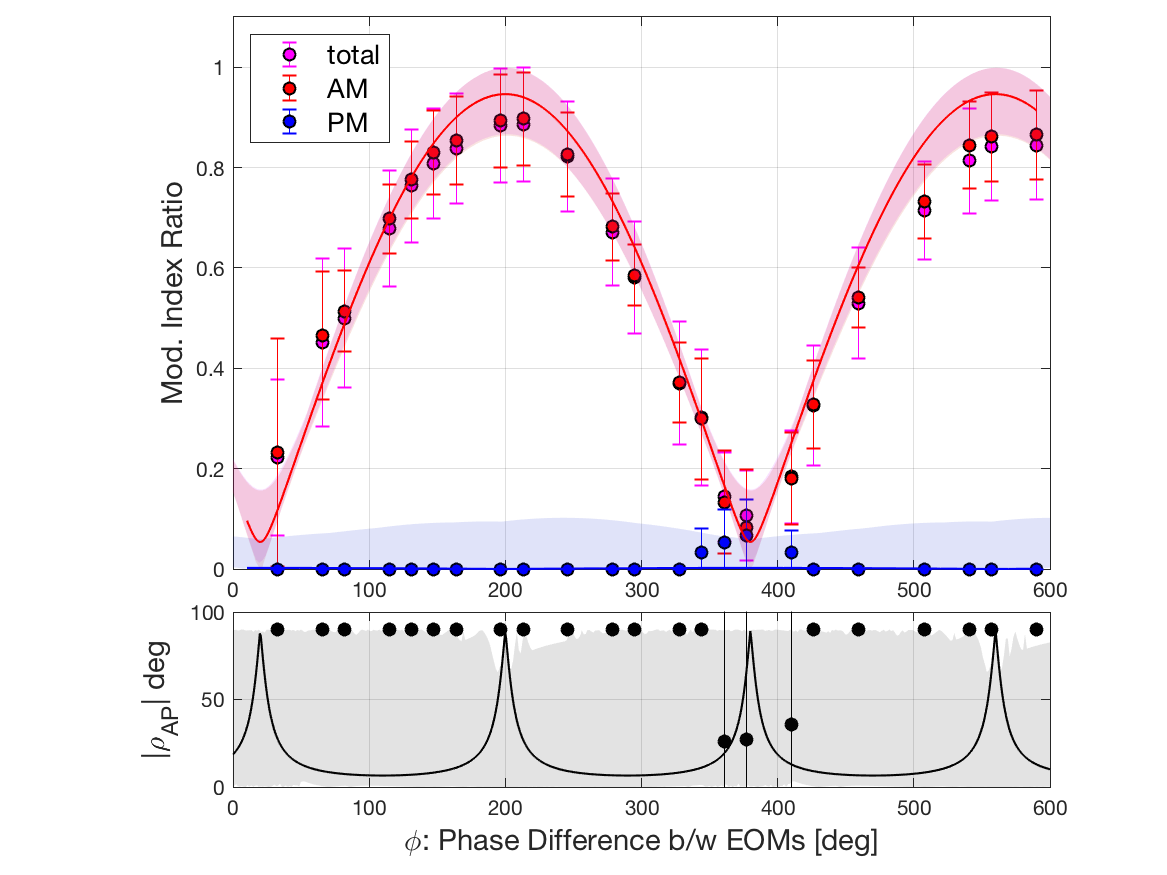}
		\subcaption{$\f{3}$}
		\label{fig:indexvsphi_b}
	\end{minipage}
	\renewcommand{\baselinestretch}{1.0}
	\caption{Dependency of modulation indices and $\rh{AP}$ on the phase difference between EOMs in the first MZI.
	Magenta, red and blue markers are the measurements for the total, AM and PM indices, respectively.
	The solid lines are from \Eref{eq:mzmoutasym} with parameters shown in \Tref{tab:indexvsphiparam}.}
	\label{fig:indexvsphi}
\end{figure}

The measurement results are shown in \Fref{fig:indexvsphi}.
The modulation indices of the PM and AM components were successfully tuned
for the sideband at $\f{1}$ as predicted by the theory
(left panel in \Fref{fig:indexvsphi}).
For $\f{3}$, the PM component was successfully eliminated by the delay line length
introduced in the second MZI and only the AM component were observed,
while the magnitude was tuned by the phase difference between the two EOMs
(right panel in \Fref{fig:indexvsphi}).
Using the Monte-Carlo method, the shaded areas are the calculated experimental uncertainties
assuming an asymmetry of the beamsplitter's reflectivity of 10\,\%, fluctuation of the amount of the modulations of 10\,\%, 
and path length deviation of 2.5\,cm for the first and 7.5\,cm for the second MZI.
The errors of the measurement were estimated from the long-term stability,
discussed in \sref{subsub:longterm}, and the systematic error of 5\,\%.
$\rh{AP}$ at $\f{3}$ is almost meaningless because of the nominally-zero PM component.\par

\begin{table}
	\caption{\label{tab:indexvsphiparam}Parameters in the measurement of the dependency
	of the modulation indices on the phase difference between EOM2a and EOM2b in the first MZI.} 
	\begin{indented}
		\item[]\begin{tabular}{@{}llll}
			\br
			Parameters & value\\
			\mr
			\multicolumn{2}{c}{Contrast}\\
			first MZI &96.82\,\% \\ 
			second MZI &94.08\,\% \\ 
			\multicolumn{2}{c}{Modulation Index at $\f{2}$}\\ 
			at EOM1 &0.1624\\ 
			locked MZM &0.0582 \\ 
			Reduction Ratio &0.3090 (theory) \\ 
			&0.3549 (measurement) \\ 
			\multicolumn{2}{c}{Modulation Index of Each EOM at $\f{1}$ and $\f{3}$}\\ 
			$\f{1}$: EOM2a &0.1572 \\ 
			$\f{1}$: EOM2b &0.1431 \\ 
			$\f{3}$: EOM2a &0.0563 \\ 
			$\f{3}$: EOM2b &0.0502 \\ 
			\br
		\end{tabular}
	\end{indented}
\end{table}

\subsubsection{Long-term stability}\label{subsub:longterm}
It is crucial from the perspective of the GW physics and astronomy
that multiple detectors in the observational network are simultaneously in an operation.
This requires each detector has a good performance on the duty cycles.
Because the modulation system plays the pivotal role in the control of the main interferometer,
the stability of the modulation system itself is also significant.
Moreover, the long-term fluctuation would degrade the AM cancellation.
This can directly cause the instability on the detector sensitivity.\par

\begin{figure}[h]
	\centering
	\includegraphics[width=10cm]{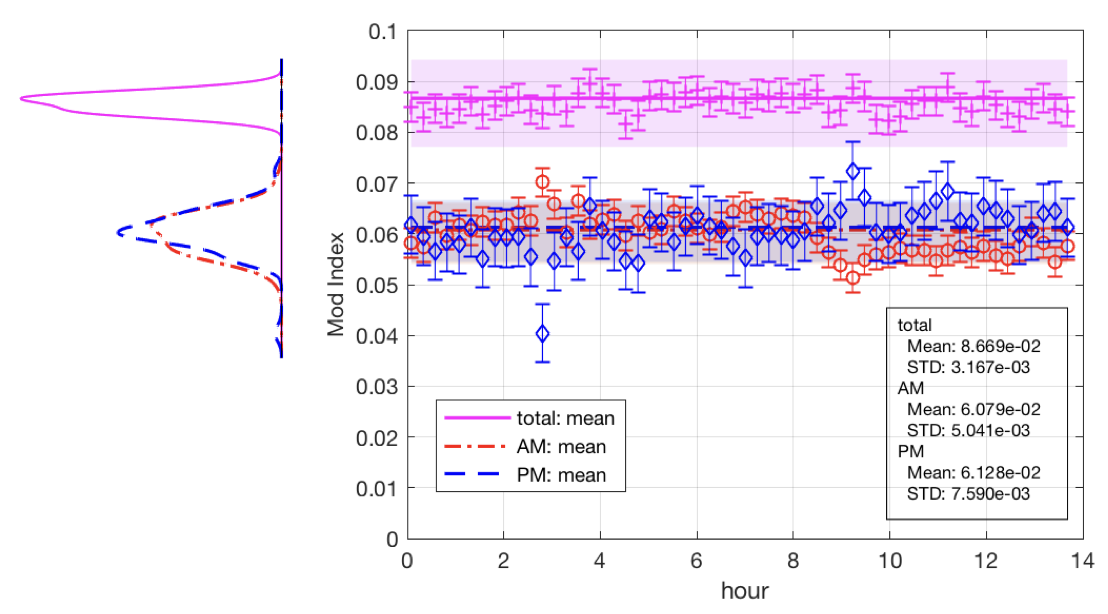}
	\renewcommand{\baselinestretch}{1.0}
	\caption{Long-term trend of the modulation index at $\f{1}$ with the phase difference of 130\,deg.
	The measurement period was 14\,hours and the data samplings were done every 5\,minutes.
	Each line is the mean value and shaded areas are the 10\,\% region.
	The error bars are estimated from the same measurement every 10\,seconds for 5\,minutes. A half of the markers are thinned out for visuality. }
	\label{fig:longtrend}
\end{figure}

To confirm a long-term performance,
fourteen hour measurements were done at multiple operating phase differences
on account of the investigation of the correlation between the standard deviation and the phase.
Here we show a long term operation with the PM and AM modulations at $\f{1}$, which simulates the case for DRSE (\Fref{fig:longtrend}).
No significant drift was observed in the 14 hour operation.
The fluctuation of the modulation indices of the PM and AM components was within the range of 10\,\% for 97\,\%
of the time during the operation.
The PM fluctuation coincides the fluctuation of the error signal for the SRC,
and the 10\,\% change can be easily compensated by a small change of the servo gain of the SRC control loop.
On the other hand, the fluctuation of the AM component corresponds to the best ratio of the AM cancellation.
The 10\,\% is not good enough because it roughly corresponds
to the reduction of the most stringent OPN requirement of -180\,dBc down to -160\,dBc.
It should be noted that the experiment was conducted with the air filters on, and the fluctuation may be reduced remarkably
in a quieter environment.\par

The results of the other measurements are listed in \Tref{tab:longtrend}.
The fluctuation of the PM at $\f{3}$ is affected by the calibration error because the PM is nominally designed to be zero.\par

\begin{table}
	\caption{\label{tab:longtrend} Parameters of the long-term measurements. Coefficient of Variation (CV) is defined by (STD)/(mean). Combined with \Fref{fig:indexvsphi}, this shows the trend: CV is larger as the magnitude of the modulation index goes smaller. The period of all the measurement is 14\,hours. The large CV of the PM at $\f{3}$ is almost from calibration error.} 
	\begin{indented}
		\item[]\begin{tabular}{@{}llllll}
			\br
			frequency & type & phase [deg]  &mean& STD& CV\\
			\mr
			f1     & total&0  &$1.45\times 10^{-1}$ &$3.61\times 10^{-3}$ & $2.49\times10^{-2}$\\ 
				    &        &131 &$8.67\times 10^{-2}$ &$3.17\times 10^{-3}$& $3.65\times 10^{-2}$ \\
				    &        &196 &$6.94\times 10^{-2}$ &$3.11\times 10^{-3}$ &$4.48\times 10^{-2}$\\
			        & AM  &0  &$7.91\times 10^{-3}$ &$5.33\times 10^{-3}$&$6.74\times 10^{-1}$ \\ 
					&        &131 &$6.08\times 10^{-2}$ &$5.04\times 10^{-3}$&$8.29\times 10^{-2}$ \\
					&        &196 &$6.41\times 10^{-2}$ &$3.28\times 10^{-3}$&$5.12\times 10^{-2}$ \\
					& PM  &0  &$1.45\times 10^{-1}$ &$3.63\times 10^{-3}$ &$2.51\times 10^{-2}$\\ 
					&        &131 &$6.13\times 10^{-2}$ &$7.59\times 10^{-3}$& $1.24\times 10^{-1}$\\
					&        &196 &$2.42\times 10^{-2}$ &$1.26\times 10^{-2}$ &$5.20\times 10^{-1}$\\
		    f3     & total&0  &$3.81\times 10^{-3}$ &$3.41\times 10^{-3}$ & $8.96\times 10^{-1}$\\ 
		    		&        &65 &$1.87\times 10^{-2}$ &$3.88\times 10^{-3}$& $2.07\times 10^{-1}$ \\
		    		&        &196 &$3.69\times 10^{-2}$ &$3.20\times 10^{-3}$ &$8.67\times 10^{-2}$\\
		    		& AM  &0  &$1.76\times 10^{-3}$ &$2.75\times 10^{-3}$&$1.56\times 10^{0}$ \\ 
		    		&        &65 &$2.10\times 10^{-2}$ &$2.79\times 10^{-3}$&$1.33\times 10^{-1}$ \\
		    		&        &196 &$3.90\times 10^{-2}$ &$3.12\times 10^{-3}$&$8.00\times 10^{-2}$ \\
		    		& PM  &0  &$3.20\times 10^{-3}$ &$3.32\times 10^{-3}$ &$1.04\times 10^{0}$\\ 
		    		&        &65 &$3.16\times 10^{-3}$ &$4.95\times 10^{-3}$& $1.56\times 10^{0}$\\
		    		&        &196 &$4.16\times 10^{-3}$ &$7.51\times 10^{-3}$ &$1.81\times 10^{0}$\\
			\br
		\end{tabular}
	\end{indented}
\end{table}

\subsubsection{Displacement noise}\label{subsub:disp}
 
The displacement noise level of each MZI was calibrated in m/$\surd$ Hz with the following equation, 
\begin{eqnarray}
\delta x =\sqrt{\frac{1}{S^{2}H^{2}}V_{ctrl}^{2}+\left(\frac{1}{1+G}\right)^{2}\left[\frac{G^{2}}{H^{2}}n_{\rm PD}^{2}+\frac{G^{2}}{SH^{2}}n_{\rm S}^{2}+A^{2}n_{\rm PZT}^{2}\right]},\label{eq:delx}
\end{eqnarray}
where $S$, $H$, $A$ and $G$ are the servo filter, the optical gain, the actuator efficiency and the open-loop transfer function, respectively; $n_{\rm PD,\hspace{0.5mm}S,\hspace{0.5mm}PZT}$ are the noise of the photodetector, the servo and the actuator, respectively; $V_{\rm ctrl}$ is the control signal. $V_{\rm ctrl}$ is considered as suppressed mechanical vibrations, {\it e.g.} excited by the seismic noise or the air flow. This is because contributions of $n_{\rm PD}$ and $n_{\rm S}$ to $V_{\rm ctrl}$ are suppressed by the factor of $\frac{1}{G}$ compared to those terms in \Eref{eq:delx} and $n_{\rm PZT}$ is expected to be smaller than the mechanical vibrations.

\begin{figure}[h]
	\begin{minipage}[h]{0.5\linewidth}
		\centering
		\includegraphics[width=7.0cm]{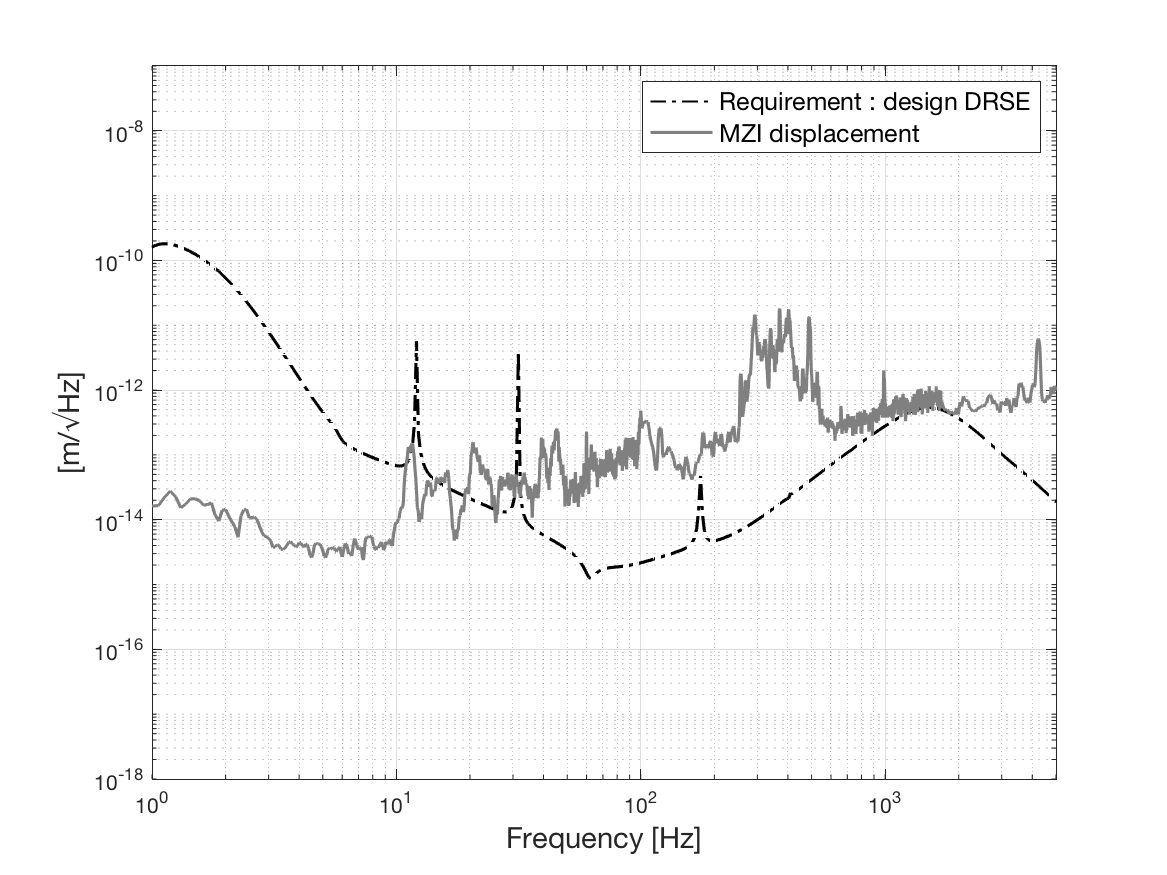}
		\subcaption{first MZI}
		\label{fig:freeinloop_a}
	\end{minipage}
	\begin{minipage}[h]{0.5\linewidth}
		\centering
		\includegraphics[width=7.0cm]{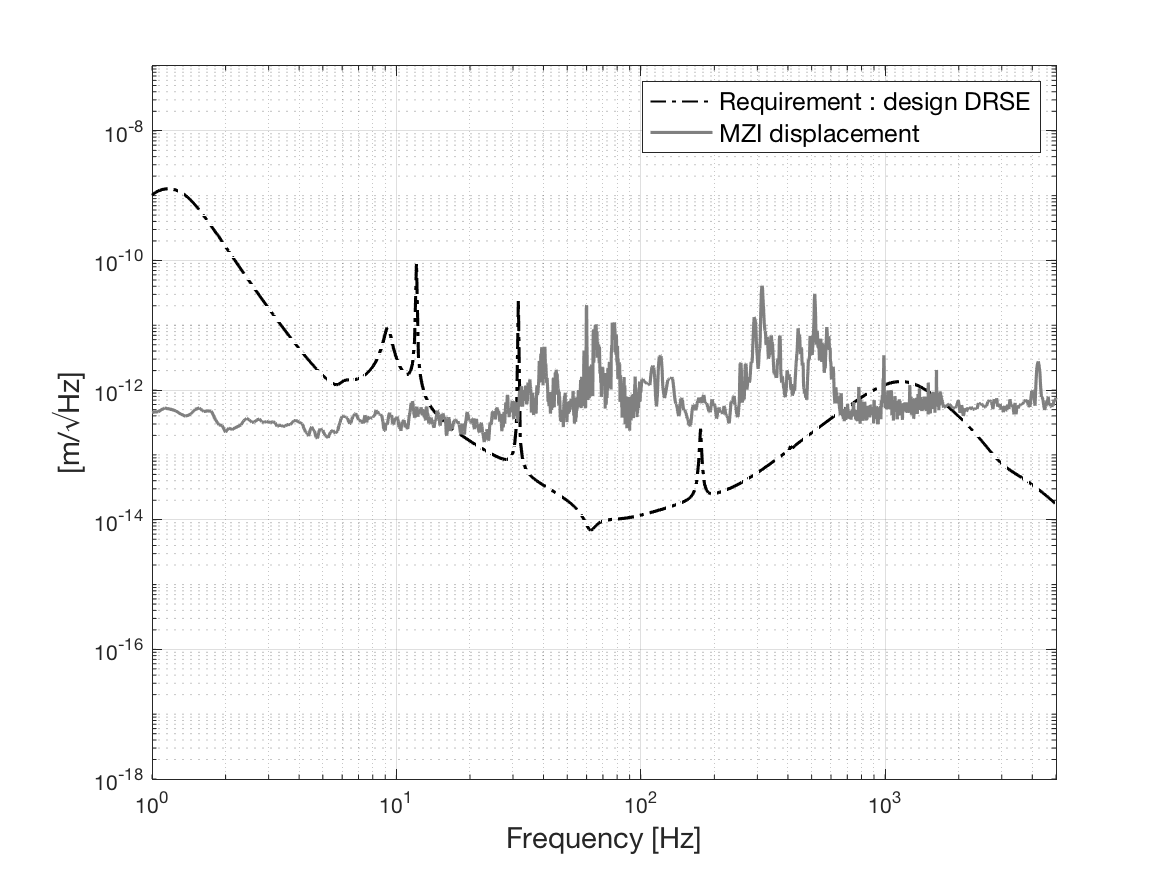}
		\subcaption{second MZI}
		\label{fig:freeinloop_b}
	\end{minipage}
	\renewcommand{\baselinestretch}{1.0}
	\caption{The measured displacements in MZIs was still above the requirement derived in \sref{sec:dispnoise} and did not currently meet the requirement. }
	\label{fig:freeinloop}
\end{figure}

\Fref{fig:freeinloop} shows the measured and calibrated displacement noise of each MZI suppressed by the feedback control loop.
The result did not meet the requirement derived in \Fref{fig:simudispreq} in Section \ref{sec:dispnoise},
from several tens to several hundreds of Hz.
Especially large peaks around several hundred Hz are seen in both of the MZIs. 

\begin{figure}[h]
	\begin{minipage}[h]{0.5\linewidth}
		\centering
		\includegraphics[width=7.0cm]{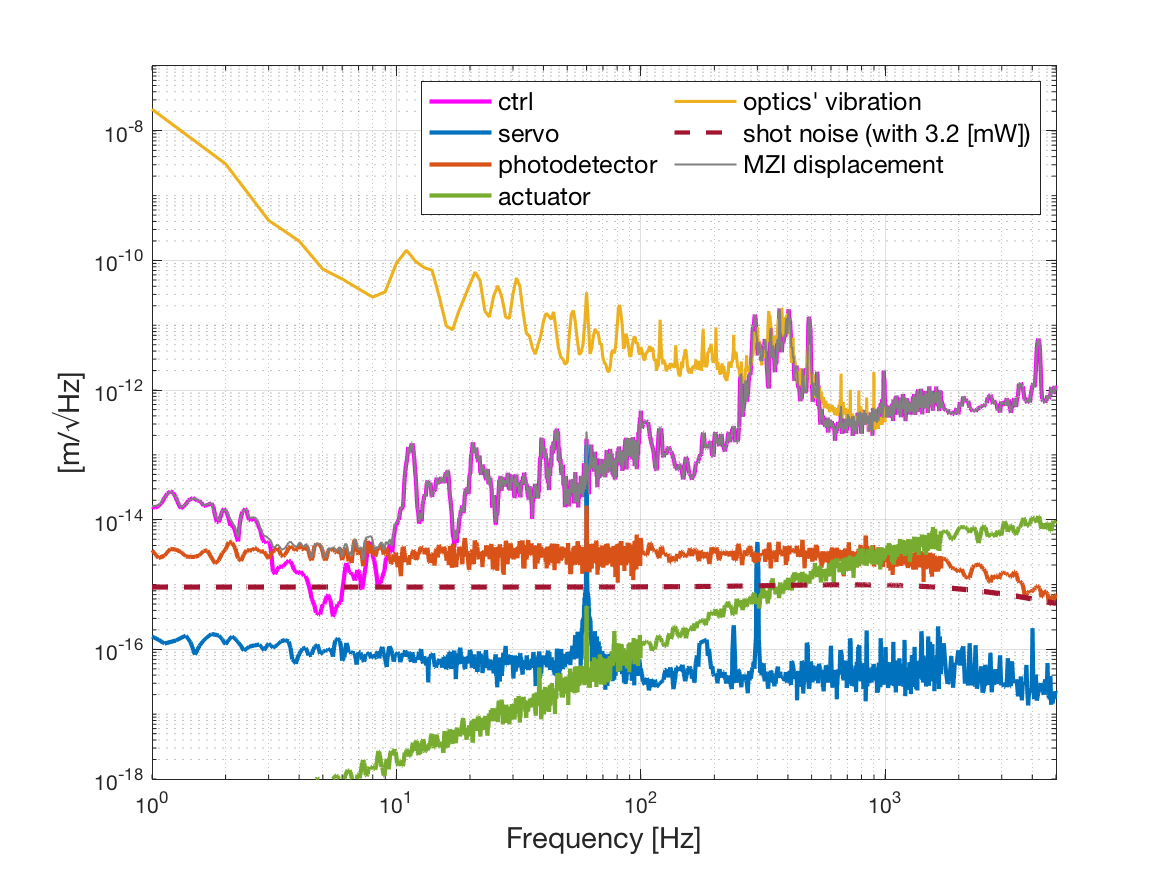}
		\subcaption{first MZI}
		\label{fig:budget_a}
	\end{minipage}
	\begin{minipage}[h]{0.5\linewidth}
		\centering
		\includegraphics[width=7.0cm]{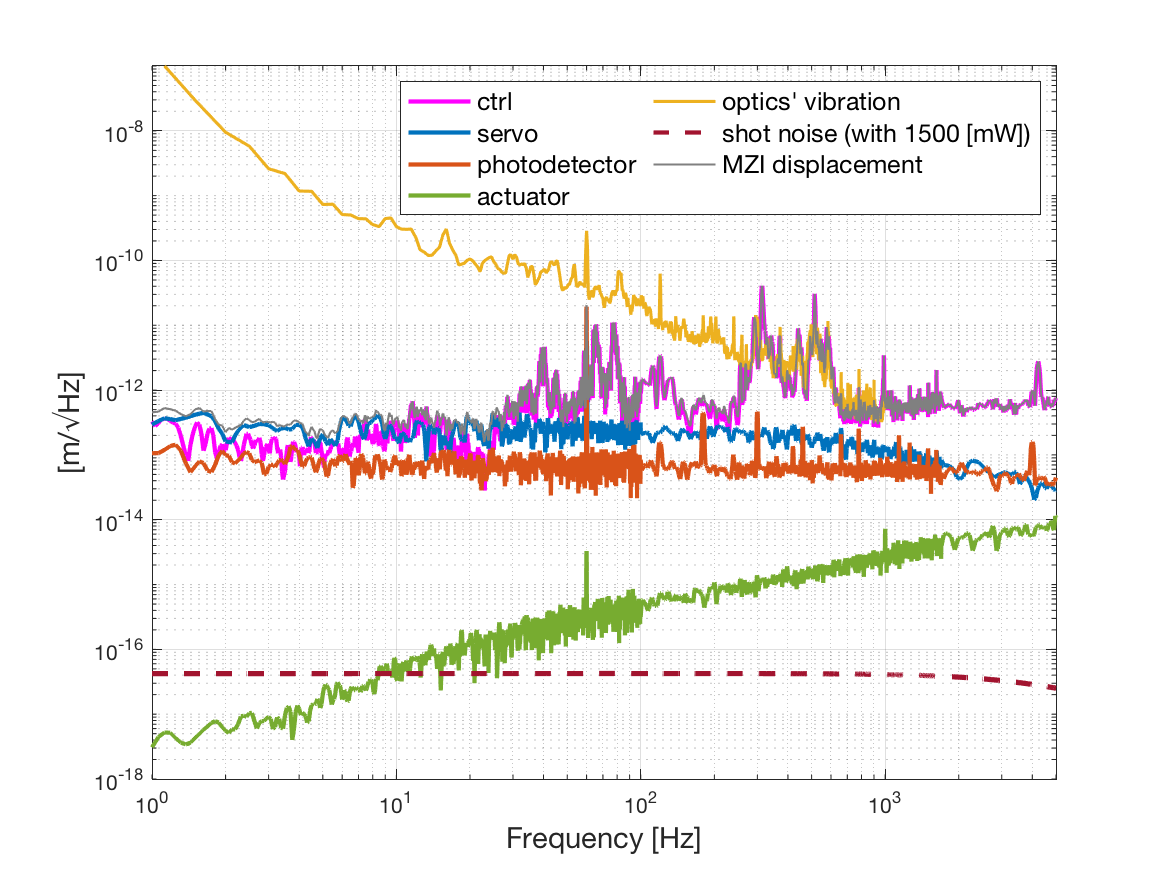}
		\subcaption{second MZI}
		\label{fig:budget_b}
	\end{minipage}
	\renewcommand{\baselinestretch}{1.0}
	\caption[Noise budget]{Noise budget: Optics vibrations are incoherently summed up for all the optics for each MZI shown in \Fref{fig:setup}. Electrical noises do not limit the displacement noise around intermediate frequency band where the requirement is toughest. On the other hand, the vibrations of optics looks greatly suppressed by the common mode rejections. }
	\label{fig:budget}
\end{figure}

To identify the origins of large peaks at several hundred Hz, noise identification efforts have made.
According to \Eref{eq:delx},
a noise budget of each MZI was made as depicted in \Fref{fig:budget}.
Noise budget is a break down plot to understand how various kinds of noise contribute to the displacement sensitivity.
In this case, the contributions of the servo noise, actuator noise,
and photodetector dark noise to the displacement were evaluated. 
In addition to the above noise sources, vibrations of the optics were studied.
A small accelerometers were attached on some MZM optics
and the vibration levels of the optics were measured.
\Fref{fig:budget} includes the accelerometer signals shown in yellow.\par

\begin{figure}[h]
	\begin{minipage}[h]{0.5\linewidth}
		\centering
		\includegraphics[width=7.0cm]{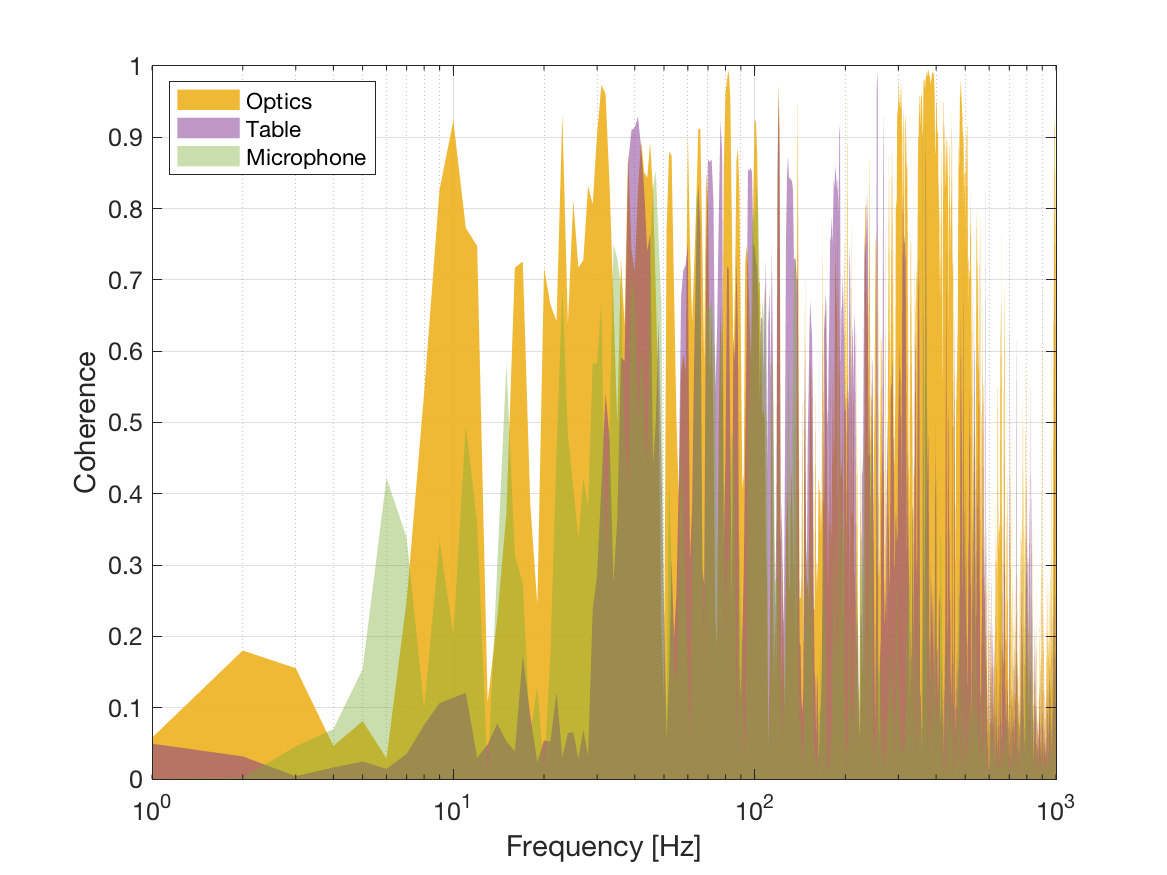}
		\subcaption{first MZI}
		\label{fig:coh_a}
	\end{minipage}
	\begin{minipage}[h]{0.5\linewidth}
		\centering
		\includegraphics[width=7.0cm]{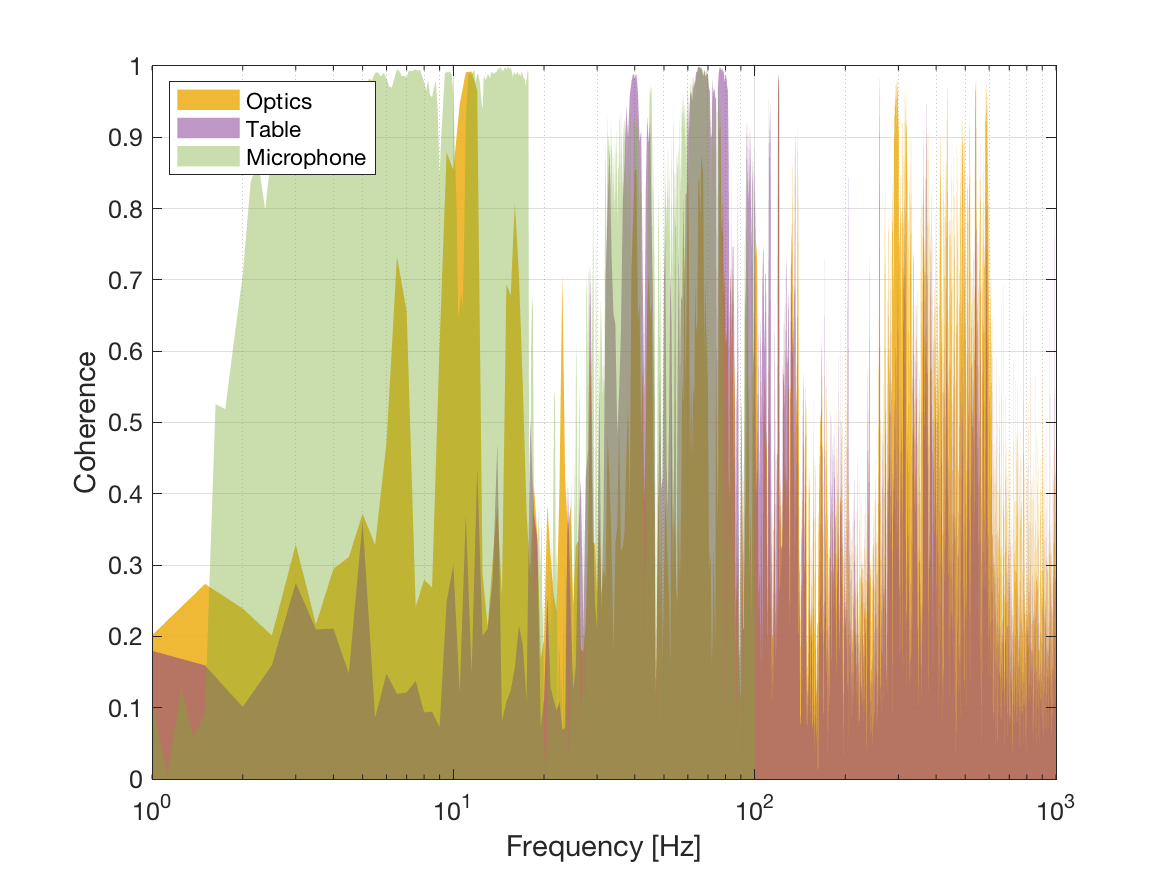}
		\subcaption{second MZI}
		\label{fig:coh_b}
	\end{minipage}
	\renewcommand{\baselinestretch}{1.0}
	\caption{Coherence function between MZIs displacement and signals of environmental monitros: microphone and accelerometers attached onto optics or table. }
	\label{fig:coh}
\end{figure}

As for the fist MZI, from 10 to 300 Hz, there are large gaps between the control signal and accelerometer signal.
Because the coherence between the two 
were broadly high in this frequency region (see, the left panel of \Fref{fig:coh}),
the mechanical vibrations were highly suppressed by common mode rejection.
On the other hand, the vibrations at higher frequencies above 200 Hz
are the mechanical resonances of the optics mounts and pedestals, and they directly contributed to the displacement sensitivity.
This is because the rejection is difficult to perform at the mechanical resonances of optics. \par

This was true for the second MZI as well, but the vibration of the optical table was more remarkable for the second MZI.
The second MZI was placed in the center of the table and much larger in size than the first MZI due to the asymmetry calculated in \Eref{eq:delayline}.
Because of these facts, the displacement sensitivity of the second MZI
could have been affected directly by the vibration mode of the table
without transferring of the vibration through the optics. See, the peaks at 40, 70 and 80 Hz
in the right panel of \Fref{fig:budget}.
Note that the displacement of the second MZI had the broad and large coherence with the signal of microphone,
see, panel (b) in \Fref{fig:coh}.
It is very likely from air fluctuations caused by clean booth filters
resulting in the refractive index fluctuations on the optical paths. \par

In summary, the present displacement noise measured in the input laser room of KAGRA did not meet the requirement. However, the dominant noise sources were well understood. The configuration which can perform great common mode rejections through the observation band, 10 to 1000 Hz, should be considered.

\par

\par

\par

\section{Discussion}\label{sec:disc}
There are several topics that remain to be studied further for the future improvements.
Two major issues are following.\par

First, the consideration of the intercorrelation of the sideband noises between the REFL port
and the AS port is important. This is because the MZM system aims at the noise cancellation
at the REFL port, while this system generates additional noise coupling to the AS port as discussed in \sref{sub:drawback}.
It is good to have more profound understandings of the intercorrelation in analytic forms and simulation.
For example, in \sref{sub:simulation}, the MZM system generating only PM components
at the frequencies of $\f{1}$ and $\f{2}$ was simply placed in the input laser system
of the detuned interferometer. However, the components of the MZM system would be tuned in an operation phase so that the noise coming from the excess AM component at the REFL port is cancelled in the DRSE configuration.
If the MZM parameters optimized for the REFL port is the worst case for the AS port,
the requirement for the displacement would be tougher.
In contrast, it is also probable that the requirement in DRSE would be relaxed.\par

Second, the measurement of the displacement noise in \sref{subsub:disp} suggests the further reduction of the mechanical vibrations and broader range of the common mode rejection are necessary. 
The followings are currently-considered future upgrades.
\begin{itemize}
	\item Change of parameters in the MZM system
	\item Put the MZM system on a base to isolate the system from the table vibration
	\item Replacement of the pedestals and holders with more rigid ones
	\item Use of the monolithic optics
	\item Put the optical paths in a vacuum
\end{itemize}
The first point mainly suggests to have a smaller size of the second MZI by enhancing the frequency of $\f{3}$ because the delay-line length is determined by \Eref{eq:delayline}. The rescaling of the second MZI can make it more insensitive to the table vibrations. These upgrades will lead the improvement of long-term stability, too.

\section{Conclusion}\label{sec:conc}
In this first R\&D research of the MZM system for DRSE, we derived the requirements for the displacement noise level, conducted a proof-of-principle experiment, and evaluated the system as the noise source. The derived requirement showed the feasibility of this system in KAGRA. The key functionality of the MZM system was first empirically demonstrated. Although the current displacement noise did not meet the requirement, we could collect informative data for the future upgrade. Towards the DRSE operation, the MZM system will be sophisticated.

\section*{Acknowledgement}
This work was supported by MEXT, JSPS Leading-edge Research Infrastructure Program, JSPS Grant-in-Aid for Specially Promoted Research 26000005, MEXT Grant-in-Aid for Scientific Research on Innovative Areas 2905: 17H06358, 17H06361 and 17H06364, JSPS Core-to-Core Program A. AdvancedResearch Networks, JSPS Grant-in-Aid for Scientific Research (S) 17H06133, the joint research program of the Institute for Cosmic Ray Research, University of Tokyo, National Research Foundation (NRF) and Computing Infrastructure Project of KISTI-GSDC in Korea, the LIGO project, and the Virgo project.

\section*{Appendix}
\subsection*{Parameters used in the simulation}
The optical parameters used for the simulation in \sref{sec:dispnoise} are summarized in \Tref{tab:simuparamifo} and \Tref{tab:simuparammod}. These parameters are the design of KAGRA.\par
\begin{table}
	\caption{\label{tab:simuparamifo}Simulation parameters (the main interferometer).}
	\begin{indented}
		\item[]\begin{tabular}{@{}llll}
			\br
			parameter & value & supplements \\
			\mr
			Arm asymmetry & 15\,ppm & \\ 
			ITMs: Power Trans&0.004 & \\ 
			ITMs: HR Loss&45\,ppm & X (Y) : +($-$) Arm asym.\\
			ETMs: Power Trans&10\,ppm & \\ 
			ETMs: HR Loss &45\,ppm & X (Y) : +($-$) Arm asym.\\ 
			PRM: Power Trans&0.10 & \\ 
			PRM: HR Loss &45\,ppm & \\
			SRM: Power Trans&0.1536-45\,ppm &\\
			SRM: HR Loss &45\,ppm & \\ 
			Arm Cavity length &3000\,m & \\
			PR Cavity length &66.5913279884\,m & \\
			SR Cavity length &(66.5913279884 - 7e5/$\f{1}$)\,m & \\
			Schnupp Asymmetry &3.32985084757\,m  &$\equiv l_{x}-l_{y}$ \\
			DARM offset &DRSE (BRSE): 1.29109 (1.29057)\,pm &\\
			Detune phase &$90-3.5=86.5$\,deg & \\ 
			Laser power at BS &780\,W &\\
			\br
		\end{tabular}
	\end{indented}
\end{table}

\begin{table}
	\caption{ \label{tab:simuparammod}Simulation parameters (Modulation \& others). }
	\begin{indented}
		\item[]\begin{tabular}{@{}llll}
			\br
			parameter & value & supplements \\
			\mr
			IMC FSR & 5626987\,Hz & \\
			$\f{1}$ & 3$\times$(MC FSR)\,Hz & \\
			$\f{2}$ & 8$\times$(MC FSR)\,Hz & \\
			$\f{3}$ & 10$\times$(MC FSR)\,Hz & \\
			mod. index at $\f{1}$ & BRSE: 0.1683\ii\,rad &\\
			mod. index at $\f{2}$ & 0.1619\ii\,rad & \\ 
			mod. index at $\f{3}$ & 0.0\,rad & \\
			Delay-line length &$0.5 \times \lambda_{3}$ & \\
			Mirror Reflectivity & 1 & \\
			BS & 1\,\% asymmetry &transmission = 0.5$\times$1.01 \\
			first MZI offset & 0\,m & $6.61\times 10^{-10}$\,m\\
			Assumed filter &low-pass filter (1\,Hz pole) & \\
			Safety factor &10 & \\
			\br
		\end{tabular}
	\end{indented}
\end{table}

\end{document}